# SHADOWING MATCHING ERRORS FOR WAVE-FRONT-LIKE SOLUTIONS

## XIAO-BIAO LIN

ABSTRACT. Consider a singularly perturbed system

$$\epsilon u_t = \epsilon^2 u_{xx} + f(u, x, \epsilon), \quad u \in \mathbb{R}^n, x \in \mathbb{R}, t \geq 0.$$

Assume that the system has a sequence of regular and internal layers occurring alternatively along the $x$-direction. These "multiple wave" solutions can formally be constructed by matched asymptotic expansions. To obtain a genuine solution, we derive a *Spatial Shadowing Lemma* which assures the existence of an exact solution that is near the formal asymptotic series provided (1) the residual errors are small in all the layers, and (2) the matching errors are small along the lateral boundaries of the adjacent layers. The method should work on some other systems like $\epsilon u_t = -(-\epsilon^2 D_{xx})^m u + \dots$.

## 1. INTRODUCTION

One of the most effective method to study wave-front-like solutions in singularly perturbed reaction-diffusion equations is the formal asymptotic method. Fife [13] studied the formation of sharp wave fronts and obtained conditions on the stability of such solutions by the formal method. Matched expansions for systems of any number of equations to any order in $\epsilon$ has also been constructed, [25]. However, the asymptotic method does not guarantee that there is an exact solution near the formal solution. Many efforts have been made to derive methods that verify the validity of formal solutions, [1, 2, 3, 5, 16, 17, 18, 20, 22, 23, 24]. We present a new method in this paper.

Although the method should work for some higher order parabolic systems, for simplicity, consider a second order singularly perturbed

*Date*: April 10, 1996.

1991 *Mathematics Subject Classification.* Primary 35B25, 35K57, Secondary 35K20, 34E15.

Research partially supported by NSFgrant DMS9002803 and DMS9205535.

system,

$$\epsilon u_t = \epsilon^2 u_{xx} + f(u, x, \epsilon), \quad u \in \mathbb{R}^n, \ x \in \mathbb{R}, \ t \geq 0. \qquad (1.1)$$

Assume that (1.1) has an infinite sequence of regular and internal layers occurring alternatively along the $x$ direction. Assuming by a matched expansion, we have constructed the following formal series for

(1) positions of moving wave fronts $\eta^\ell(t, \epsilon) = \sum_0^m \epsilon^j \eta_j^\ell(t)$, $\ell \in \mathbb{Z}$;

(2) solutions in the $\ell$th regular layer $u^{R\ell}(x, t, \epsilon) = \sum_0^m \epsilon^j u_j^{R\ell}(x, t)$;

(3) solutions in the $\ell$th internal layer $u^{S\ell}(\xi, t, \epsilon) = \sum_0^m \epsilon^j u_j^{S\ell}(\xi, t)$,   $(1.2)$

where $\xi = (x - \eta^\ell(t, \epsilon))/\epsilon$.

Here the superscript "R" and "S" stand for regular and singular (internal) layers.

Although our method can be applied to systems with various kinds of boundary conditions. To simplify the illustration, assume that the nonlinear term is periodic in $x$, and we are looking for solutions that are periodic in $x$. Assume that the formal expansions are periodic in $\ell$, compatible with the period of $f$. That is, there exists a constant $x_p > 0$ and an integer $\ell_p > 0$, such that the following holds:

**Periodicity Hypotheses**

1. $f(u, x + x_p, \epsilon) = f(u, x, \epsilon)$;
2. $\eta^{\ell+\ell_p}(t, \epsilon) = \eta^\ell(t, \epsilon) + x_p$;
3. $u^{R(\ell+\ell_p)}(x, t, \epsilon) = u^{R\ell}(x - x_p, t, \epsilon)$;
4. $u^{S(\ell+\ell_p)}(\xi, t, \epsilon) = u^{S\ell}(\xi, t, \epsilon)$.

Due to the periodicity of $f$ and the formal solutions, all the estimates in this paper are uniformly valid with respect to the layer index $\ell$, and this will not be repeated in the paper. The periodicity plays no other rule apart from mentioned above.

Through out this paper, let $0 < \beta < 1$ be a fixed constant. Let the width of the internal layer be $2\epsilon^\beta$. Let

$$y^{2\ell}(t) = \eta^\ell(t, \epsilon) + \epsilon^\beta,$$

$$y^{2\ell-1}(t) = \eta^\ell(t, \epsilon) - \epsilon^\beta.$$

A family of curves $\Gamma^i = \{(x, t) : x = y^i(t), t \geq 0\}$, $i \in \mathbb{Z}$, divides the domain $\{x \in \mathbb{R}\} \times \{t \geq 0\}$ into infinitely many strips, $\Sigma^i = \{(x, t) : y^{i-1}(t) \leq x \leq y^i(t), t \geq 0\}$, $i \in \mathbb{Z}$. $\Sigma^i$ is an internal (or regular) layer if $i = 2\ell$ (or $i = 2\ell - 1$).

A doubly infinite sequence will be denoted by $\{h^i\}$, with the norm $|\{h^i\}|_{E^i} = \sup_i\{|h^i|_{E^i}\}$ if each $h^i$ is in a Banach space $E^i$. A sequence of functions $\{w^i(x,t)\}$, each defined in $\Sigma^i$, is called a *formal approximation* subordinate to the partition, if the residual error in $\Sigma^i$,

$$- g^i = \epsilon w_t^i - \epsilon^2 w_{xx}^i - f(w^i, x, \epsilon), \tag{1.3}$$

and the jump error at the boundary $\Gamma^i$,

$$- \delta^i = \begin{pmatrix} I \\ \epsilon D_x \end{pmatrix}(w^i - w^{i+1}), \tag{1.4}$$

are small. We also call the piecewise smooth function $w$ which is equal to $w^i$ in $\Sigma^i$ a formal approximation. In singular perturbation problems, the formal approximation can naturally be provided by matched expansions. For $(x,t) \in \Sigma^i$,

$$w^i(x,t,\epsilon) = \begin{cases} u^{R\ell}(x,t,\epsilon), & i = 2\ell - 1, \\ u^{S\ell}(\frac{x - \eta^\ell(t,\epsilon)}{\epsilon}, t, \epsilon), & i = 2\ell. \end{cases} \tag{1.5}$$

We also assume that $y^i(t)$ and $w^i(t)$ are defined for all $t \geq 0$ and approaches $y^i(\infty)$ and $w^i(\infty)$ as $t \to \infty$. (Our method can show the existence of exact solutions in finite time if $y^i$ and $w^i$ exist in finite time.)

Under certain conditions, we want to show that the smallness of residual and jump errors implies that for each initial condition $u(x,0,\epsilon)$ near $w(x,0,\epsilon)$, there is a unique exact solution $u$ near $w$. To this end, let $u = w^i + u^i$ in $\Sigma^i$. The correction terms $\{u^i\}$ satisfy nonlinear equations with the forcing terms $g^i$, as in (1.3), in $\Sigma^i$,

$$\epsilon u_t^i = \epsilon u_{xx}^i + f(w^i + u^i, x, \epsilon) - f(w^i, x, \epsilon) + g^i, \tag{1.6}$$

and jump conditions $\delta^i$, as in (1.4), at $\Gamma^i$,

$$\begin{pmatrix} I \\ \epsilon D_x \end{pmatrix}(u^i - u^{i+1}) = \delta^i. \tag{1.7}$$

The small solutions $\{u^i\}$ can be solved by linearization and contraction mappings in certain Banach spaces. The process is much like a shadowing lemma argument used in [23]. However, since the jumps do not occur in the temporal direction, we can not directly use the normal *Temporal Shadowing Lemma* as in that paper. A new *Spatial Shadowing Lemma* is needed to correct jump errors along the spatial direction.

Several coordinate systems are used in this paper. We denote the original coordinate system, as used in (1.1), by $\mathfrak{R}(0)$. Other coordinates will be defined through $\mathfrak{R}(0)$ and will be denoted by $\mathfrak{R}(1)$, $\mathfrak{R}(2)$, etc.

We will denote a two dimensional region $\Sigma^i \cap \{\mathbb{R} \times I\}$, where $I$ is a time interval, by $\Sigma^i \cap I$ for simplicity.

We now illustrates basic ideas used in this paper. Consider a coordinate change in the region $\Sigma^i \cap [0, \epsilon \Delta \tau]$,

$$\mathfrak{R}(0) \to \mathfrak{R}^i(1) : \Sigma^i \cap [0, \epsilon \Delta \tau] \ni (x,t) \to (\xi, \tau) \in \Omega^i \times I_\tau,$$
$$y = x - \Xi(x, t, \epsilon),$$
$$\xi = [y - (y^{i-1} + y^i)]/\epsilon, \quad \tau = t/\epsilon.$$

Here $y = x - \Xi(x, t, \epsilon)$ is a near identity change of coordinates that straightens $\Gamma^{i-1}$ & $\Gamma^i$ so that the functions $y^{i-1}$ & $y^i$ are time independent. $\Omega^i = \{-L^i(\epsilon) \leq \xi \leq L^i(\epsilon)\}$, $L^i(\epsilon) = (y^i - y^{i-1})/2\epsilon$ and $I_\tau = [0, \Delta \tau]$.

Using the coordinate system $\mathfrak{R}^i(1)$ in the region $\Omega^i \times I_\tau$, we linearize system (1.6) at a fixed time $\tau = 0$. We are led to

$$u^i_\tau = u^i_{\xi\xi} + V^i(\xi) u^i_\xi + A^i(\xi) u + \mathcal{N}^i(u^i, \xi, \tau, \epsilon) + g^i, \qquad (1.8)$$

where $V^i$ is a scalar multiplier, and $\mathcal{N}^i$ is a small term. Eqn (1.8) is to be solved with an initial condition

$$u^i_0(\xi) = u(\xi, 0, \epsilon) - w^i(\xi, 0, \epsilon). \qquad (1.9)$$

The change of variable can also be made such that it does not affect the jump condition (1.7) at $\Gamma^i$. For illustration, let $\mathcal{N}^i = 0$. Assuming that the coefficients of Eqn. (1.8) can be extended to $\xi \in \mathbb{R}$, as well as $u^i_0$ and $g^i$. Using the variation of constant formula in a suitable Banach space, (1.8) can be solved for $\xi \in \mathbb{R}$ with nonzero $u^i_0$ and $g^i$ but no boundary conditions. We then only have to treat (1.8) with $u^i_0 = 0$, $g^i = 0$ and some jump boundary conditions.

Define $H_0^{0.75 \times 0.25}(I_\tau) = H_0^{0.75}(I_\tau) \times H_0^{0.25}(I_\tau)$. Let $u^i \in H^{2,1}(\Omega^i \times I_\tau)$, from the Trace Theorem [28], the mapping $\xi \to \begin{pmatrix} u^i(\xi, \cdot) \\ u^i_\xi(\xi, \cdot) \end{pmatrix}$, $\Omega^i \to H_0^{0.75 \times 0.25}(I_\tau)$ is continuous. However, Eqn (1.8) does not generate a flow in $H_0^{0.75 \times 0.25}(I_\tau)$. Following the idea of [21, 30], we want to find subspaces $W^s(\xi) \oplus W^u(\xi) = H_0^{0.75 \times 0.25}(I_\tau)$ such that the linear homogeneous equation associate to (1.8) can be solved for $\xi > \xi_0$ if the boundary value $\begin{pmatrix} u^i(\xi_0) \\ u^i_\xi(\xi_0) \end{pmatrix} = \begin{pmatrix} \phi_1 \\ \phi_2 \end{pmatrix} \in W^s(\xi_0)$ (or $\xi < \xi_0$ if $\begin{pmatrix} \phi_1 \\ \phi_2 \end{pmatrix} \in W^u(\xi_0)$). We also want to show that these solutions decay as $\xi$ moves away from $\xi_0$. Then the idea of *Temporal Shadowing Lemma* can be used to solve systems (1.8) and (1.7).

The dichotomy splitting can easily be achieved by applying the Fourier–Laplace transform to (1.8). Using $\wedge$ for images of Laplace transforms, we have a first order system, which will be called a dual system associated to (1.8).

$$D_\xi \begin{pmatrix} \hat{u}^i(s) \\ \hat{v}^i(s) \end{pmatrix} = \begin{pmatrix} 0 & I \\ sI - A^i(\xi) & -V^i(\xi)I \end{pmatrix} \begin{pmatrix} \hat{u}^i(s) \\ \hat{v}^i(s) \end{pmatrix} + \begin{pmatrix} 0 \\ \hat{\mathcal{N}}^i + \hat{g}^i \end{pmatrix}, \tag{1.10}$$

Equation (1.10) is an ODE in a Banach space of homomorphic functions (Hardy–Lebesgue classes). It can be shown that in certain region of $s$, (1.10) has an exponential dichotomy. Then the inverse Laplace transform gives the desired dichotomy in $H_0^{0.75 \times 0.25}(I_\tau)$.

We have treated the problem in a short time interval. In the coordinates $\mathfrak{R}(0)$, let $I_j = [j\Delta t, (j+1)\Delta t]$, $0 \leq j \leq r-1$, and $I_\infty = [t_f, \infty)$. Here $\Delta t = \epsilon \Delta \tau$ and $t_f = r\Delta t$. The constant $t_f$ has to be so large and $\Delta t > 0$ so small that the variations of $y^i(t)$ and $w^i(t)$ with respect to $t$ are small in each time interval. We can then use a near identity change of coordinates to straighten the boundary $\Gamma^i$, $i \in \mathbb{Z}$, and use a linearization at a frozen time with very small errors. Nonlinear systems like (1.8) can be solved recursively in time intervals $I_j$, $0 \leq j \leq r-1$, and $I_\infty$, using the values of $u^i$ at the end of the $I_j$ as the initial data for $I_{j+1}$. We remark that the number of intervals $r+1 \to \infty$ as $\epsilon \to 0$. The main difficulty in this paper is to control the growth of $u^i$ as $r$ gets larger.

The outline of this paper is as follows. We state our main result, the *Spatial Shadowing Lemma* in §2. The proof of that lemma is also included there, but should be read after §6, since Theorems 5.1, 6.1 and 6.2 are used in the proof. The theory of dichotomy splittings is developed in §3 through several technical lemmas. A useful property on Evans function, due to Gardner and Jones [19], is used to prove Lemma 3.10. Readers who want to know the main flow of proofs can skip §3 on the first reading, but come back for those lemmas when they are quoted. In §4, we introduce the change of coordinates that is used to straighten the curve $\Gamma^i$. We also study a system of linear equations with jump conditions along their common boundaries (Lemma 4.1). The result there is the main tool to handle jump boundary conditions in the next two sections. In §5, we treat equations in the final interval $I_\infty$. The result obtained there also tells us how small the upper bound of $|u_{exact} - w|$ must be at time $t = t_f$ in order to construct exact solutions in $I_\infty$. In §6, we treat equations on each time interval $I_j$, $0 \leq j \leq r-1$. The error on each $I_j$ has to be small so that the accumulation on all $r$ intervals yields a small error at $t = t_f$. In §7, the *Spatial Shadowing Lemma* is applied to the singularly perturbed system (1.1). We briefly review the construction of matched asymptotic expansions and formal approximations for (1.1) and the hypotheses used for such constructions, [25]. We show that these formal approximations satisfy the hypotheses of the Spatial Shadowing Lemma, as stated in §2. A precise relation between eigenvalues and wave speeds of internal layers is given in Lemma 7.1, which may be of some independent interest.

We have concentrated on global wave-front-like solutions in this paper. However, our method can treat short time solutions with much weaker hypotheses. These local solutions are discussed in [27]. The result obtained there applies to various general systems including Cahn-Hilliard eqautions and viscous profile for conservation laws, which do not satify our hypotheses in §2.

## 2. Main Result

Let $\Omega$ and $I$ be space and time intervals respectively. Define the following Banach spaces and norms.

$H^{2,1}(\Omega \times I) = \{u(x,t) : \Omega \times I \to \mathbb{R}^n \mid u, \ u_{xx} \ \text{and} \ u_t \in L^2(\Omega \times I; \mathbb{R}^n)\}.$
$|u|_{H^{2,1}(\Omega \times I)} = |u|_{L^2} + |u_{xx}|_{L^2} + |u_t|_{L^2}.$
$H^r(I) = W^{r,2}(I; \mathbb{R}^n), \ r \geq 0, \ \text{is the usual Soblev space}.$
$H^{r \times s}(I) = H^r(I) \times H^s(I), \quad r \geq 0, \ s \geq 0.$

For a constant $\gamma \in \mathbb{R}$, let

$\quad L^2(\Omega \times \mathbb{R}^+, \gamma) = \{u : \Omega \times \mathbb{R}^+ \to \mathbb{R}^n \mid e^{-\gamma t}u \in L^2(\Omega \times \mathbb{R}^+)\}.$
$\quad |u|_{L^2(\Omega \times \mathbb{R}^+, \gamma)} = |e^{-\gamma t}u|_{L^2(\Omega \times \mathbb{R}^+)}.$
$\quad H^{2,1}(\Omega \times \mathbb{R}^+, \gamma) = \{u : \Omega \times \mathbb{R}^+ \to \mathbb{R}^n \mid e^{-\gamma t}u \in H^{2,1}(\Omega \times \mathbb{R}^+)\}.$
$\quad |u|_{H^{2,1}(\Omega \times \mathbb{R}^+, \gamma)} = |e^{-\gamma t}u|_{H^{2,1}(\Omega \times \mathbb{R}^+)}.$

For a constant $\gamma < 0$, let

$X(\Omega \times \mathbb{R}^+, \gamma) = \{u : u = u_1 + u_2, \ u_1 \in L^2(\Omega), \ u_2 \in L^2(\Omega \times \mathbb{R}^+, \gamma)\}.$
$|u|_{X(\gamma)} = |u_1|_{L^2(\Omega)} + |u_2|_{L^2(\Omega \times \mathbb{R}^+, \gamma)}.$
$X^{2,1}(\Omega \times \mathbb{R}^+, \gamma) = \{u : u = u_1 + u_2, \ u_1 \in H^2(\Omega), \ u_2 \in H^{2,1}(\Omega \times \mathbb{R}^+, \gamma)\}.$
$|u|_{X^{2,1}(\gamma)} = |u_1|_{H^2(\Omega)} + |u_2|_{H^{2,1}(\Omega \times \mathbb{R}^+, \gamma)}.$

It can be verified that for $u \in X(\Omega \times \mathbb{R}^+, \gamma)$ or $X^{2,1}(\Omega \times \mathbb{R}^+, \gamma)$, the decomposition $u = u_1 + u_2$ is unique. We often use simplified notations, for example $X^{2,1}(\gamma)$, instead of $X^{2,1}(\Omega \times \mathbb{R}^+, \gamma)$.

Let $\Sigma = \{(x,t) : y^1(t) < x < y^2(t), \ t \geq 0\}$, where $y^1$, $y^2$ are smooth functions of $t$, be a two dimensional region unbouded in the $t$ direction. We say that $u \in L^2(\Sigma, \gamma)$, $H^{2,1}(\Sigma, \gamma)$, $X(\Sigma, \gamma)$ or $X^{2,1}(\Sigma, \gamma)$, etc., if $u$ is the restriction of a function $U \in L^2(\mathbb{R} \times \mathbb{R}^+, \gamma)$, etc., to the domain

$\Sigma$. The norms are defined by

$$|u|_{L^2(\Sigma,\gamma)} = \inf\{|U|_{L^2(\mathbb{R}\times\mathbb{R}^+,\gamma)}\},$$

$$|u|_{X^{2,1}(\Sigma,\gamma)} = \inf\{|U|_{X^{2,1}(\mathbb{R}\times\mathbb{R}^+,\gamma)}\}.$$

Let

$$X^k(\gamma) = \{u \mid u = u_1 + u_2,\ u_1 \in \mathbb{R}^n,\ u_2 \in H^k(\gamma)\},\ \gamma < 0.$$
$$X^{k_1 \times k_2}(\gamma) = X^{k_1}(\gamma) \times X^{k_2}(\gamma), \quad k_1 \geq 0,\ k_2 \geq 0.$$

If $I$ is a finite time interval, similar definitions can be given to spaces $L^2(\Omega \times I, \gamma)$, $H^{2,1}(\Omega \times I, \gamma)$, $H^{2,1}(\Sigma \cap I, \gamma)$, but not $X(\Omega \times I, \gamma)$ or $X^k(\gamma)$,

As in §1, let $\{w^i\}$ be a formal approximation for (1.1). Let $\{\eta^\ell\}$ be a formal approximation for the wave fronts. We extend the domain of $w^i$ from $\Sigma^i$ to $\mathbb{R} \times \mathbb{R}^+$ by letting

$$w^i(x,t,\epsilon) = \begin{cases} w^i(y^{i-1}(t),t,\epsilon), & \text{if } x < y^{i-1}(t), \\ w^i(y^i(t),t,\epsilon), & \text{if } x > y^i(t). \end{cases}$$

Assume

**H 1.** There exist $C$, $\bar\gamma > 0$ such that for all small $\epsilon$ and $i, \ell \in \mathbb{Z}$,

$$|w^i(x,t,\epsilon) - w^i(x,\infty,\epsilon)| \leq Ce^{-\bar\gamma t}, \quad (x,t) \in \mathbb{R}^2.$$

$$|\eta^\ell(t,\epsilon) - \eta^\ell(\infty,\epsilon)| + |D_t\eta^\ell(t,\epsilon)| \leq Ce^{-\bar\gamma t}, \quad t \in \mathbb{R}^+.$$

Here $w^i(x,\infty,\epsilon) = \lim_{t\to\infty} w^i(x,t,\epsilon)$, etc.

**H 2.** There exists $\bar\sigma > 0$ such that in each regular layer $\Sigma^i$, $i = 2\ell - 1$,

$$\operatorname{Re}\sigma\{f_u(w^i(x,t,\epsilon),x,\epsilon)\} \leq -\bar\sigma$$

uniformly for all $(x,t) \in \Sigma^i$, $i \in \mathbb{Z}$ and small $\epsilon > 0$.

**H 3.** In an internal layer, as $\xi \to \pm\infty$ and $\epsilon \to 0$, both $w^i$ and $\partial w^i/\partial\xi$ approach the corresponding values of $w^{i+1}$ or $w^{i-1}$ at common boundaries. More precisely, if $i = 2\ell$, then for any $\mu > 0$, there exist $N$, $\epsilon_0 > 0$ such that $\epsilon_0^{\beta-1} > N$, and for $0 < \epsilon < \epsilon_0$, $t \geq 0$,

$$|w^i(\xi,t,\epsilon) - w^{i-1}(y^{i-1}(t),t,\epsilon)| \leq \mu, \quad \text{for } -\epsilon^{\beta-1} \leq \xi \leq -N,$$
$$|w^i(\xi,t,\epsilon) - w^{i+1}(y^i(t),t,\epsilon)| \leq \mu, \quad \text{for } \epsilon^{\beta-1} \geq \xi \geq N.$$

Here the function $w^i$ is expressed in the stretched variable $\xi = (x - \eta^\ell(t,\epsilon))/\epsilon$.

Let $\tilde\xi = \tilde\xi(\xi)$ be a function of $\xi$ such that

$$\tilde\xi = \begin{cases} \tilde\xi, & \text{for } |\xi| \leq \epsilon^{\beta-1}, \\ -\epsilon^{\beta-1}, & \text{for } \xi < -\epsilon^{\beta-1}, \\ \epsilon^{\beta-1}, & \text{for } \xi > \epsilon^{\beta-1}. \end{cases}$$

For each $t \geq 0$, $i = 2\ell$, consider the operator $\mathcal{A}^i(t) : L^2(\mathbb{R}) \to L^2(\mathbb{R})$,

$$\mathcal{A}^i(t)u = u_{\xi\xi} + D_t \eta^\ell(t, \epsilon)u_\xi + f_u(w^i(\tilde{\xi}, t, \epsilon), \eta^\ell(t, \epsilon) + \epsilon\tilde{\xi}, \epsilon)u.$$

**H 4.** $\mathcal{A}^i(t)$, $i = 2\ell$, $t \geq 0$, has a simple eigenvalue $\lambda^i(t, \epsilon) = \epsilon\lambda_0^i(t) + O(\epsilon^2)$. The rest of the spectrum is contained in $\{\mathrm{Re}\lambda \leq -\bar{\sigma}\}$, $\bar{\sigma}$ as in **H2**. Moreover, for the limiting operator $\mathcal{A}^i(\infty)$, we have,

$$\lambda_0^i(\infty) \leq \overline{\lambda}_0 < 0, \quad \text{uniformly for all } i = 2\ell.$$

We now introduce a coordinate system $\mathfrak{R}^i(2)$ for the region $\Sigma^i$.

$$\mathfrak{R}(0) \to \mathfrak{R}^i(2) \,:\, \Sigma^i \ni (x, t) \to (\xi, \tau) \in \tilde{\Sigma}^i,$$
$$\xi = [x - \tfrac{1}{2}(y^{i-1}(t) + y^i(t))]/\epsilon, \quad \tau = t/\epsilon.$$

Here the image of $\Sigma^i$ is denoted by $\tilde{\Sigma}^i$.

In the new variables the residual and jump errors are

$$- g^i = w_\tau^i - w_{\xi\xi}^i - f^i, \tag{2.1}$$

$$-\delta^i(\tau) = \begin{pmatrix} I \\ D_\xi \end{pmatrix} (w^i([y^i - y^{i-1}]/2\epsilon, \tau, \epsilon) - w^{i+1}([y^i - y^{i+1}]/2\epsilon, \tau, \epsilon)). \tag{2.2}$$

Here $f = f(w^i, \tfrac{1}{2}(y^{i-1}+y^i)+\epsilon\xi, \epsilon)$ with $w^i = w^i(\xi, \tau, \epsilon)$, and $y^i = y^i(\epsilon\tau)$.

The main result of this paper is the following

**Spatial Shadowing Lemma .** *Let $\{\bar{w}^i\}$ be a formal approximation of solutions for (1.1). Assume that the Hypotheses **H1**–**H4** are satisfied. Then there exist positive constants $j_0$, $J_2$, $\epsilon_0$ and a negative constant $\gamma \leq -C_3\epsilon$ satisfying the following properties. Assume that $\{w^i\}$ is a formal approximation near $\{\bar{w}^i\}$, with*

$$|w^i - \bar{w}^i|_{X^{2,1}(\tilde{\Sigma}^i, \gamma)} \leq C_1 \epsilon^{j_1}, \quad i \in \mathbb{Z}, \tag{2.3}$$

*for some $C_1 > 0$ and $j_1 \geq 1$. Assume that for the approximation $\{w^i\}$, we have*

$$|g^i|_{X(\tilde{\Sigma}^i, \gamma)} + |\delta^i|_{X^{0.75 \times 0.25}(\mathbb{R}^+, \gamma)} \leq C_2 \epsilon^{j_2}, \quad j_2 \geq j_0. \tag{2.4}$$

*Then for $0 < \epsilon < \epsilon_0$, to any locally $H^1$ function $u_0$ with $|u_0 - w^i(0)|_{H^1(\tilde{\Sigma}^i \cap \{\tau=0\})} \leq C_2 \epsilon^{j_2}$, there exists a unique exact solution to (1.1) that satisfies $u_{exact}(x, 0) = u_0(x)$, and*

$$|u_{exact} - \bar{w}^i|_{X^{2,1}(\tilde{\Sigma}^i, \gamma))} = O(\epsilon^{j_3}), \quad i \in \mathbb{Z}.$$

*Here $j_3 = \min\{j_1, j_2 - J_2\}$, $J_2 > 0$ is a constant that does not depends on $\epsilon$. All the norms in this lemma are expressed by the coordinate system $\mathfrak{R}^i(2)$.*

**Remark** We explain why the second approximation $\{w^i\}$ is needed in the lemma. Suppose that $\{\bar{w}^i\}$ comes from a matched expansion of order $\epsilon^m$. The constant $j_0$ is determined by $\{\bar{w}^i\}$. $j_0$ is stable with respect to small perturbations. Unfortunately it is not guaranteed that the residual and jump errors for $\{\bar{w}^i\}$ are bounded by $C_2\epsilon^{j_0}$. The formal approximation $\{w^i\}$, on the other hand, can be obtained by adding higher order expansions to $\{\bar{w}^i\}$. It will satisfy (2.3), therefore, share the same $j_0$ with $\{\bar{w}^i\}$. Also its residual and jump errors, will be bounded by $C_2\epsilon^{j_2}$ for any desired $j_2 > 0$ if sufficiently many terms are added.

Since we know how to compute formal approximations to any desired accuracy [25], the specific values of $j_0$, $J_2$ are not crucial. This was first observed in [4]. The constant $j_1$ is important since $j_3 = j_1$ if $j_2$ is large. To compute $j_1$, we can take $\{w^i\}$ as an expansion of order $\epsilon^{m+1}$, for higher order $\{w^i\}$ yields the same $j_1$.

*Proof of the Spatial Shadowing Lemma.* The proof is based on Theorems 5.1, 6.1 and 6.2.

Let $\Delta\tau$, $\Delta t$, $t_f$, $r > 0$ be the constants as in Theorem 6.2. In the coordinates $\mathfrak{R}(0)$, let $I_j = [j\Delta t, (j+1)\Delta t]$, $0 \leq j \leq r-1$ and $I_\infty = [t_f, \infty)$. A partition of the time axis

$$\mathbb{R}^+ = \bigcup_{j=0}^{r-1} I_j \bigcup I_\infty,$$

has been constructed in §6. Let $\gamma < 0$ be as in Theorem 5.1.

At each $\bar{t} = j\Delta t$, $0 \leq j \leq r-1$, or $\bar{t} = \infty$, with the $X^{2,1}$ norm evaluated by the coordinates $\mathfrak{R}^i(2)$,

$$|w^i(\bar{t}) - \bar{w}^i(\bar{t})|_\infty \leq C|w^i - \bar{w}^i|_{X^{2,1}(\tilde{\Sigma}^i, \gamma)} \leq CC_1\epsilon^{j_1} \leq C\epsilon.$$

Recall that the coefficients of the linear equations in §5 and §6 are obtained by linearizing at $\bar{t}$. Thus the linearization around $\{w^i(\bar{t})\}$ and $\{\bar{w}^i(\bar{t})\}$ are $\epsilon$–close. The constant $J_1$ in Theorem 6.2 depends on the coefficients of the linearization in a complicated way, but is stable with respect to small perturbations of the coefficients. Therefore, there exists $\epsilon_0 > 0$ such that the same $J_1$ is shared by all the $\{w^i\}$ satisfying (2.3). Let now $j_0 = r_1 + J_1 + 1.5$, where $r_1 > 1.5$.

Since $j_2 > j_0 = r_1 + J_1 + 1.5$, all the terms in (2.4) are bounded by $o(\epsilon^{r_1+1.5+J_1})$. We now use the coordinates $\mathfrak{R}^i(1)$, as in §4, in each region $\Sigma^i \cap I_j$, $0 \leq j \leq r-1$ or $j = \infty$. Recall that in the coordinates $\mathfrak{R}^i(1)$, $\Sigma^i \cap I_j$ maps to $\Omega_j^i \times I_\tau$ where $I_\tau = [0, \Delta\tau]$, and $\Sigma^i \cap I_\infty$ maps to $\Omega_\infty^i \times \mathbb{R}^+$. One can verify that the residual and jump errors are still of $O(\epsilon^{j_2}) = o(\epsilon^{r_1+J_1+1.5})$ in the coordinates $\mathfrak{R}^i(1)$. After all, in the region

$\Sigma^i \cap I_j$, the two coordinate changes

$$\mathfrak{R}(0) \to \mathfrak{R}^i(1),$$
$$\mathfrak{R}(0) \to \mathfrak{R}^i(2).$$

only differ by a near identity mapping $y = x - \Xi(x, t, \epsilon)$. Therefore, in the finite time intervals $I_j$, Theorems 6.1 and 6.2 can be applied to obtain a unique exact solution, for each initial data $u_0$ satisfying $|u_0 - w^i(0)|_{H^1(\Omega_0^i)} \leq C_2 \epsilon^{j_2}$. According to Theorem 6.2, the accumulation error on all these intervals are small so that the solution in the final interval $I_\infty$ is guaranteed by Theorem 5.1. Denote the solutions by $w^i + u_j^i$, $0 \leq j \leq r-1$, in $\Sigma^i \cap I_j$ and $w^i + u_\infty^i$ in $\Sigma^i \cap I_\infty$. The solutions satisfy

$$\begin{aligned} |u_j^i|_{H^{2,1}(\Omega_j^i \times I_\tau)} &\leq C\epsilon^{j_2 - J_1 - 1.5}, &&\text{in } \Omega_j^i \times I_\tau, \\ |u_\infty^i|_{X^{2,1}(\Omega_\infty^i \times \mathbb{R}^+, \gamma)} &\leq C\epsilon^{j_2 - J_1 - 2}, &&\text{in } \Omega_\infty^i \times \mathbb{R}^+. \end{aligned}$$

It is easy to verify that, with the left side evaluated by the coordinates $\mathfrak{R}^i(2)$ and the right side by $\mathfrak{R}^i(1)$,

$$|u^i|_{X^{2,1}(\tilde{\Sigma}^i, \gamma)} \leq Ce^{|\gamma|r\Delta\tau} \Big[ \sum_0^{r-1} |u_j^i|_{H^{2,1}(\Omega_j^i \times I_\tau)} + |u_\infty^i|_{X^{2,1}(\Omega_\infty^i \times \mathbb{R}^+, \gamma)} \Big].$$

Since $r \approx \log(1/(C_0\epsilon^2))/(\epsilon\bar{\gamma}\Delta\tau)$ and $|\gamma| \leq C_3\epsilon$, cf. §6, we have

$$e^{|\gamma|r\Delta\tau} \leq Ce^{C_3 \log(1/(C_0\epsilon^2))/\bar{\gamma}} \leq C[1/(C_0\epsilon^2)]^{C_3/\bar{\gamma}} \leq C\epsilon^{-J_3}.$$

The number of the terms in the summation is $r = o(\epsilon^{-1.5})$ if $\epsilon$ is small. Thus,

$$|u^i|_{X^{2,1}(\tilde{\Sigma}^i, \gamma)} \leq C\epsilon^{j_2 - J_1 - J_3 - 3}.$$

Let now $J_2 = J_1 + J_3 + 3$. The proof of the lemma has been completed. $\square$

## 3. BASIC LEMMAS

A function $f(s)$ is in the Hardy-Lebesgue class $\mathcal{H}(\gamma)$, $\gamma \in \mathbb{R}$, if
(i) $f(s)$ is analytic in $\text{Re}(s) > \gamma$;
(ii) $\{\sup_{\sigma > \gamma} (\int_{-\infty}^\infty |f(\sigma + i\omega)|^2 d\omega)\}^{1/2} < \infty$.
$\mathcal{H}(\gamma)$ is a Banach space with the norm defined by the left side of (ii). Based on the Paley-Wiener Theorem, [32], if $e^{-\gamma t}f(t) \in L^2(\mathbb{R}^+)$, then $\hat{f}(s) \in \mathcal{H}(\gamma)$, vice versa.

For $k \geq 0$ and $\gamma \in \mathbb{R}$, define a Banach space

$$\mathcal{H}^k(\gamma) = \{u(s) \mid u(s) \text{ and } (s-\gamma)^k u(s) \in \mathcal{H}(\gamma)\},$$
$$|u|_{\mathcal{H}^k(\gamma)} = |u|_{\mathcal{H}(\gamma)} + |(s-\gamma)^k u|_{\mathcal{H}(\gamma)}.$$

For any $\gamma \in \mathbb{R}$, $k \geq 0$, there exists $C = C(\gamma, k)$ such that

$$C^{-1}(1 + |s|^k) \leq 1 + |s - \gamma|^k \leq C(1 + |s|^k).$$

Therefore an equivalent norm for $\mathcal{H}^k(\gamma)$ is

$$|u|^2_{\mathcal{H}^k(\gamma)} = \sup_{\sigma > \gamma} \int_{-\infty}^{\infty} |u(\sigma + i\omega)|^2 (1 + |\sigma + i\omega|^{2k}) d\omega.$$

It can be shown that if $e^{-\gamma t} f(t) \in H_0^k(\mathbb{R}^+)$, then $\hat{f}(s) \in \mathcal{H}^k(\gamma)$.

Let $\mathcal{H}^{k_1 \times k_2}(\gamma) = \mathcal{H}^{k_1}(\gamma) \times \mathcal{H}^{k_2}(\gamma)$. Let $H_0^{r \times s}(\mathbb{R}^+) = H_0^r(\mathbb{R}^+) \times H_0^s(\mathbb{R}^+)$. Clearly, $(e^{-\gamma t} u, e^{-\gamma t} v) \in H_0^{r \times s}(\mathbb{R}^+)$ if and only if $(\hat{u}(s), \hat{v}(s)) \in \mathcal{H}^{k_1 \times k_2}(\gamma)$. Define a $s$-dependent norm in $\mathbb{R}^{2n}$. For $(u, v)^\tau \in \mathbb{R}^{2n}$,

$$\left| \begin{pmatrix} u \\ v \end{pmatrix} \right|_{E^{k_1 \times k_2}(s)} = (1 + |s|^{k_1})|u| + (1 + |s|^{k_2})|v|.$$

An equivalent norm for $(u, v)^\tau \in \mathcal{H}^{k_1 \times k_2}(\gamma)$ is

$$\sup_{\sigma > \gamma} [\int_{-\infty}^{\infty} |(u, v)^\tau|^2_{E^{k_1 \times k_2}(\sigma + i\omega)} d\omega]^{1/2}. \tag{3.1}$$

Consider a linear equation

$$u_t = u_{\xi\xi} + V(\xi) u_\xi + A(\xi) u + f(t). \tag{3.2}$$

Here $A(\xi)$ is a $n \times n$ matrix, continuous in $\xi$, and $V(\xi)$ is a continuous scalar function in $\xi$. Applying the Laplace transform to (3.2), we have

$$s\hat{u} = \hat{u}_{\xi\xi} + V(\xi) u_\xi + A(\xi) u + \hat{f}(s). \tag{3.3}$$

Convert it into a first order system

$$D_\xi \begin{pmatrix} \hat{u} \\ \hat{v} \end{pmatrix} = \begin{pmatrix} 0 & I \\ sI - A(\xi) & -V(\xi)I \end{pmatrix} \begin{pmatrix} \hat{u} \\ \hat{v} \end{pmatrix} + \begin{pmatrix} 0 \\ \hat{f} \end{pmatrix}. \tag{3.4}$$

Let $T(\xi, \zeta; s)$ be the principal matrix solution for (3.4) with $s$ as a parameter. Let $\mathcal{S}$ be a subset of the complex plane and $\Omega \subset \mathbb{R}$ be an interval. We say (3.4) has an exponential dichotomy in $\mathbb{R}^{2n}$ for $s \in \mathcal{S}$ and $\xi \in \Omega$, if for each $s \in \mathcal{S}$, (3.4) has an exponential dichotomy in $\mathbb{R}^{2n}$ for $\xi \in \mathbb{R}$. The projections $P_s(\xi, s) + P_u(\xi, s) = I$ in $\mathbb{R}^{2n}$ are analytic in $s$ and continuous in $\xi$. With the $s$-dependent constants $K(s)$, $\alpha(s) > 0$, we have

$$T(\xi, \zeta; s) P_s(\zeta, s) = P_s(\xi, s) T(\xi, \zeta; s), \quad \xi \geq \zeta,$$

$$|T(\xi, \zeta; s) P_s(\zeta, s)|_{\mathbb{R}^{2n}} \leq K(s) e^{-\alpha(s)|\xi - \zeta|}, \quad \xi \geq \zeta,$$

$$|T(\xi, \zeta; s) P_u(\zeta, s)|_{\mathbb{R}^{2n}} \leq K(s) e^{-\alpha(s)|\xi - \zeta|}, \quad \xi \leq \zeta.$$

We say (3.4) has an exponential dichotomy in $E^{k_1 \times k_2}(s)$ for $s \in \mathcal{S}$ and $\xi \in \Omega$, if there exist projections $P_s(\xi, s) + P_u(\xi, s) = I$ in $E^{k_1 \times k_2}(s)$, analytic in $s$ and continuous in $\xi$, and constants $K$, $\alpha > 0$, independent of $s \in \mathcal{S}$, such that

$$T(\xi, \zeta; s)P_s(\zeta, s) = P_s(\xi, s)T(\xi, \zeta; s), \quad \xi \geq \zeta,$$

$$|T(\xi, \zeta; s)P_s(\zeta, s)|_{E^{k_1 \times k_2}(s)} \leq Ke^{-\alpha(\sqrt{|s|}+1)|\xi-\zeta|}, \quad \xi \geq \zeta,$$

$$|T(\xi, \zeta; s)P_u(\zeta, s)|_{E^{k_1 \times k_2}(s)} \leq Ke^{-\alpha(\sqrt{|s|}+1)|\xi-\zeta|}, \quad \xi \leq \zeta.$$

We say (3.4) has an exponential dichotomy in $\mathcal{H}^{k_1 \times k_2}(\gamma)$ for $\xi \in \Omega$, if there exist projections $P_s(\xi) + P_u(\xi) = I$ in $\mathcal{H}^{k_1 \times k_2}(\gamma)$, continuous in $\xi$, and constants $K$, $\alpha > 0$, independent of $s \in \mathcal{S}$. And if by specifying $(\hat{u}, \hat{v})$ at $\zeta \in \Omega$, solving (3.4) and denoting the solution map by $T(\xi, \zeta)$, we have

(a) $T(\xi, \zeta) : \mathcal{R}P_s(\zeta) \to \mathcal{R}P_s(\xi)$ is defined and continuous for $\xi \geq \zeta$;

(b) $T(\xi, \zeta) : \mathcal{R}P_u(\zeta) \to \mathcal{R}P_u(\xi)$ is defined and continuous for $\xi \leq \zeta$;

(c) $\begin{aligned} &|T(\xi, \zeta)P_s(\zeta)|_{\mathcal{H}^{k_1 \times k_2}(\gamma)} \leq Ke^{-\alpha|\xi-\zeta|}, \quad \xi \geq \zeta, \\ &|T(\xi, \zeta)P_u(\zeta)|_{\mathcal{H}^{k_1 \times k_2}(\gamma)} \leq Ke^{-\alpha|\xi-\zeta|}, \quad \xi \leq \zeta. \end{aligned}$

**Lemma 3.1.** *Assume that (3.4) has an exponential dichotomy in $E^{0.75 \times 0.25}(s)$ for $\mathrm{Re}(s) \geq \gamma$ and $\xi \in [a, b]$. Then*

*(1) (3.4) has an exponential dichotomy in $\mathcal{H}^{0.75 \times 0.25}(\gamma)$ with the same projections derived from those in $E^{0.75 \times 0.25}(s)$ and the same constants $K$, $\alpha$.*

*(2) Assume that $\sup_\xi\{|A(\xi)| + |V(\xi)|\} \leq M$. Let $\phi \in \mathcal{H}^{0.75 \times 0.25}(\gamma)$, $(u, v)^\tau(\xi, s) = \mathcal{L}^{-1}(T(\xi, a; s)P_s(a, s)\phi(s))$. Then $u \in H^{2,1}([a, b] \times \mathbb{R}^+, \gamma)$ and is a solution to (3.2) with $f = 0$. Also*

$$|u|_{H^{2,1}(\gamma)} \leq C|\phi|_{\mathcal{H}^{0.75 \times 0.25}(\gamma)}.$$

*Similar result also holds for $(u, v)^\tau = \mathcal{L}^{-1}(T(\xi, b; s)P_u(b, s)\phi(s))$.*

*Proof.* Let $\phi \in \mathcal{H}^{0.75 \times 0.25}(\gamma)$. Let $w(\xi, s) = T(\xi, \zeta; s)P_s(\zeta, s)\phi(s)$, $\xi \geq \zeta \geq a$ with $P_s(\zeta, s)$ being the projection associate to the exponential dichotomy in the space $E^{0.75 \times 0.25}(s)$. The function $w$ is clearly analytic for $\mathrm{Re}(s) > \gamma$ and satisfies (3.4), with $\hat{f} = 0$.

Let $s = \sigma + i\omega$. To show (1), for each $\sigma > \gamma$, $\xi \geq \zeta$, using (3.1),

$$\int_{-\infty}^{\infty} |w(\xi, s)|^2_{E^{0.75 \times 0.25}(s)} d\omega \leq \int_{-\infty}^{\infty} K^2 e^{-2\alpha(1+|s|^{0.5})(\xi-\zeta)}|\phi|^2_{E^{0.75 \times 0.25}(s)} d\omega$$
$$\leq K^2 e^{-2\alpha(\xi-\zeta)}|\phi|^2_{\mathcal{H}^{0.75 \times 0.25}(\gamma)}.$$

This proves that $w \in \mathcal{H}^{0.75 \times 0.25}(\gamma)$ with the desired decay estimate. It is also clear that $w \in \mathcal{R}P_s$, since the projection is derived from that in $E^{0.75 \times 0.25}(s)$, therefore, is commutative with the solution map.

This proves half of (1). The other half can be proved by considering $w(\xi, s) = T(\xi, \zeta; s)P_u(\zeta, s)\phi(s), \xi \leq \zeta \leq b$.

To show (2), let $(\hat{u}, \hat{v})^\tau = w(\xi, s) = T(\xi, a; s)P_s(a, s)\phi(s), \xi \geq a$, and $s = \sigma + i\omega$. For a fixed $\sigma > \gamma$, we want to show $|\hat{u}(\xi, s)|$ and $|s\hat{u}(\xi, s)|$ are in $L^2([a, b] \times \mathbb{R})$, as functions of $(\xi, \omega)$.

$$\int_{-\infty}^{\infty}\int_a^b (1 + |s|^2)|\hat{u}|^2 d\xi d\omega \leq C \int_{-\infty}^{\infty}\int_a^b (1 + |s|^{0.5})|w|^2_{E^{0.75 \times 0.25}(s)} d\xi d\omega$$

$$\leq CK^2 \int_{-\infty}^{\infty}\int_a^b (1 + |s|^{0.5})e^{-2\alpha(1 + |s|^{0.5})(\xi - a)}|\phi|^2_{E^{0.75 \times 0.25}(s)} d\xi d\omega$$

$$\leq \frac{CK^2}{2\alpha}|\phi|^2_{\mathcal{H}^{0.75 \times 0.25}(\gamma)}.$$

By the inverse Fourier-Laplace transform, we have

$$|e^{-\sigma t}u|_{L^2([a,b]\times\mathbb{R}^+)} + |e^{-\sigma t}u_t|_{L^2([a,b]\times\mathbb{R}^+)} \leq C|\phi|_{\mathcal{H}^{0.75 \times 0.25}(\gamma)}.$$

Letting $\sigma \to \gamma$, we found that $u, u_t \in L^2([a, b] \times \mathbb{R}^+, \gamma)$. Similarly we can show that $|v|_{L^2(\gamma)} = |u_\xi|_{L^2(\gamma)} \leq C|\phi|_{\mathcal{H}^{0.75 \times 0.25}(\gamma)}$. From the equation (3.2) itself and $\sup_\xi\{|A| + |V|\} \leq M$, $|u_{\xi\xi}|_{L^2(\gamma)} \leq C|\phi|_{\mathcal{H}^{0.75 \times 0.25}(\gamma)}$. This completes the proof of (2). □

A consequence of Lemma 3.1, is that if (3.4) has an exponential dichotomy in $E^{0.75 \times 0.25}(s)$ for $\text{Re}(s) > \gamma$ and $\xi \in [a, b]$, then (3.2) has an exponential dichotomy in $H_0^{0.75 \times 0.25}(\gamma)$ for $\xi \in [a, b]$. That is, for any $\xi_0 \in [a, b]$ and $(u^0, v^0) \in H_0^{0.75 \times 0.25}(\gamma)$, there exists a decomposition $(u^0, v^0) = (u_s^0, v_s^0) + (u_u^0, v_u^0)$ such that (3.2), with $f = 0$, can be solved in $[\xi_0, b]$ with $(u(\xi_0), u_\xi(\xi_0)) = (u_s^0, v_s^0)$. It can also be solved in $[a, \xi_0]$ with $(u(\xi_0), u_\xi(\xi_0)) = (u_u^0, v_u^0)$. Moreover, for some $K, \alpha > 0$,

$$\left|\begin{pmatrix} u(\xi) \\ u_\xi(\xi) \end{pmatrix}\right|_{H_0^{k_1 \times k_2}(\gamma)} \leq Ke^{-\alpha|\xi - \xi_0|}\left|\begin{pmatrix} u^0 \\ v^0 \end{pmatrix}\right|_{H_0^{k_1 \times k_2}(\gamma)}$$

In the next lemma, we study $u_\tau = u_{\xi\xi}, u(\xi, 0) = 0$ and its Laplace transform in some detail. Consider

$$\hat{u}_{\xi\xi} - s\hat{u} = 0, \quad s \in \mathfrak{S}(M) = \{|\arg(s)| \leq 2\pi/3, |s| \geq M\}, \tag{3.5}$$

where $M \geq 1$ is a constant. Converting into a first order system, we have

$$\hat{u}_\xi = \hat{v}, \tag{3.6}$$
$$\hat{v}_\xi = s\hat{u}.$$

The eigenvalues are $\lambda = \pm\sqrt{s}$ and the eigenspace for each $\lambda$ is $n$-dimensional. We choose the major branch of $\sqrt{s}$. Thus $|\arg(\sqrt{s})| \leq$ $\pi/3$, $\mathrm{Re}\sqrt{s} \geq \sqrt{|s|}/2$. Eqn (3.6) has an exponential dichotomy in $\xi \in \mathbb{R}$. Moreover, the decay rates are $e^{-\sqrt{|s|}|\xi|/2}$. One can easily verify that

$$P_s(s) \begin{pmatrix} \hat{u} \\ \hat{v} \end{pmatrix} = \frac{1}{2} \begin{pmatrix} \hat{u} + \hat{v}/\sqrt{s} \\ \sqrt{s}\hat{u} + \hat{v} \end{pmatrix},$$

$$P_u(s) \begin{pmatrix} \hat{u} \\ \hat{v} \end{pmatrix} = \frac{1}{2} \begin{pmatrix} \hat{u} - \hat{v}/\sqrt{s} \\ -\sqrt{s}\hat{u} + \hat{v} \end{pmatrix},$$

where $P_s$ and $P_u$ are spectral projections associated to the eigenvalues $-\sqrt{s}$ and $\sqrt{s}$ respectively.

For any $k \geq 0$, using the $s$-dependent norm in $\mathbb{R}^{2n}$, we have

$$\left| P_s(s) \begin{pmatrix} \hat{u} \\ \hat{v} \end{pmatrix} \right|_{E^{(k+0.5) \times k}(s)} = \frac{1}{2} \left( |\hat{u} + \hat{v}/\sqrt{s}|(1 + |s|^{k+0.5}) + |\sqrt{s}\hat{u} + \hat{v}|(1 + |s|^k) \right)$$

$$\leq \frac{3}{2}(|\hat{u}|(1 + |s|^{k+0.5}) + |\hat{v}|(1 + |s|^k)).$$

Here we have used the fact $|s| \geq 1$. Thus $|P_s(s)|_{E^{(k+0.5) \times k}(s)} \leq 1.5$. Similarly we can show that $|P_u(s)|_{E^{(k+0.5) \times k}(s)} \leq 1.5$. Observe that for $|s| \geq 1$, $\xi \geq 0$, we have $e^{-\sqrt{|s|}\xi/2} \leq e^{-(1+\sqrt{|s|})\xi/4}$. We have thus proved the following

**Lemma 3.2.** *Eqn (3.6) has an exponential dichotomy in $E^{(k+0.5) \times k}(s)$, $k \geq 0$, for $s \in \mathfrak{S}(1)$ and $\xi \in \mathbb{R}$, with $K = 1.5$ and $\alpha = 0.25$.*

**Remark** The angle $\frac{2\pi}{3}$ in the definition of the sector can be replaced by any $\frac{\pi}{2} < \theta < \pi$. However, the constants $K$ and $\alpha$ will be different.

We shall often use the well-known *Roughness of Exponential Dichotomy* Theorem. See Coppel [6] for a proof for $I = \mathbb{R}+$, $\mathbb{R}^-$ or $\mathbb{R}$. The proof extends to the case $I$ being a finite interval. Below, we generalize the lemma to the space $E^{(k+0.5) \times k}(s)$, $k \geq 0$.

**Lemma 3.3.** *Assume that an ODE in $\mathbb{R}^{2n}$,*

$$D_\xi Y = A(\xi, s)Y$$

*has an exponential dichotomy in $E^{(k+0.5) \times k}(s)$, $k \geq 0$ for $\xi \in I$, and $s \in \mathcal{S}$, where $\mathcal{S} \subset \mathbb{C}$, and $I$ can be $\mathbb{R}$, $\mathbb{R}^+$, $\mathbb{R}^-$, or a finite interval $[a, b]$. Let the projections be $P_s(\xi, s) + P_u(\xi, s) = I$, and the constants be $K$, $\alpha > 0$. Then there exists $c_0 > 0$ such that if $\delta \stackrel{def}{=} \sup_{\xi \in I} |B(\xi)| < c_0 \alpha (1 + M^{0.5})$ for some $M \geq 0$, then*

$$D_\xi Y = [A(t) + B(t)]Y,$$

*also has an exponential dichotomy in* $E^{(k+0.5)\times k}(s)$ *with* $\xi \in I$ *and* $s \in \mathcal{S} \cap \{|s| \geq M\}$. *Let the projections for the new system be* $\tilde{P}_s(\xi, s) + \tilde{P}_u(\xi, s) = I$ *and the constants* $\tilde{K}$, $\tilde{\alpha} > 0$. *Then there exist* $c_1$, $c_2$, $c_3 > 0$ *such that* $|P_s(\xi, s) - \tilde{P}_s(\xi, s)|_{E^{(k+0.5)\times k}(s)} \leq c_1 \delta/(\alpha(1+M^{0.5}))$, $|\alpha - \tilde{\alpha}| \leq c_2 \delta/(1+M^{0.5})$, *and* $\tilde{K} \leq c_3$.

*Proof.* Following the idea of [6], or [26], Theorem 3.5, we first obtain projections to the stable and unstable spaces for the perturbed system in $E^{(k+0.5)\times k}(s)$. Solutions on these spaces decay like $e^{-(\alpha(1+|s|^{0.5})-c_2\delta)|\xi-\zeta|}$, if the initial point is $\zeta$. However,

$$\alpha(1+|s|^{0.5}) - c_2\delta > \alpha(1+|s|^{0.5}) - \frac{c_2\delta}{1+M^{0.5}}(1+|s|^{0.5}).$$

This explains that the decay rate has the form $(\alpha - \dfrac{c_2\delta}{1+M^{0.5}})(1+|s|^{0.5})$ $\qquad\blacksquare$

**Lemma 3.4.** *Consider (3.4) with* $\sup_\xi\{|A(\xi)|+|V(\xi)|\} \leq a < \infty$. *Let* $k \geq 0$ *be a constant. Then there exists* $M \geq 1$, *depending on* $a$, *such that (3.4) has an exponential dichotomy in* $E^{(k+0.5)\times k}(s)$ *for* $s \in \mathfrak{S}(M)$ *and* $\xi \in \mathbb{R}$.

*Proof.* From Lemma 3.2, system (3.6) has an exponential dichotomy with $K = 1.5$ and $\alpha = 0.25$ in the region $s \in \mathfrak{S}(M)$, $M \geq 1$. We then use Lemma 3.3. We can choose $M$ larger so that $\delta = a < 0.25c_0(1+M^{0.5})$. The result follows from Lemma 3.3. $\qquad\blacksquare$

**Corollary 3.5.** *(1) Consider (3.4) with the same hypotheses as in Lemma 3.4. Let* $\mathcal{S} \subset \mathbb{C}$ *be a closed subset. Assume that* $\mathcal{S} \cap \{|s| \geq M\} \subset \mathfrak{S}(M)$ *for some* $M > 0$. *Assume that (3.4) has an exponential dichotomy in* $\mathbb{R}^{2n}$ *for* $s \in \mathcal{S}$ *and* $\xi \in \mathbb{R}$. *Then (3.4) has an exponential dichotomy in* $E^{(k+0.5)\times k}(s)$, $k \geq 0$, *for* $s \in \mathcal{S}$ *and* $\xi \in \mathbb{R}$.

*(2) Let* $\sigma_0 \in \mathbb{R}$. *Assume that (3.4) has an exponential dichotomy in* $\mathbb{R}^{2n}$ *for* $Re(s) \geq \sigma_0$ *and* $\xi \in \mathbb{R}$. *Then it has an exponential dichotomy in* $E^{(k+0.5)\times k}(s)$, $k \geq 0$, *for* $Re(s) \geq \sigma_0$ *and* $\xi \in \mathbb{R}$.

*Proof.* Let $M \geq 1$ be the constant as in Lemma 3.4. We can choose $M$ larger so that $\mathcal{S} \cap \{|s| \geq M\} \subset \mathfrak{S}(M)$. By Lemma 3.4, (3.4) has an exponential dichotomy in $E^{(k+0.5)\times k}(s)$ for $s \in \mathcal{S} \cap \mathfrak{S}(M)$ with constants $K_1$, $\alpha_1$.

In the compact subset $\mathcal{S} \cap \{|s| \leq M\}$, (3.4) has an exponential dichotomy in $\mathbb{R}^{2n}$ with constants $K_2$, $\alpha_2$. The norm of $E^{(k+0.5)\times k}(s)$ is equivalent to the norm of $\mathbb{R}^{2n}$ uniformly with respect to $s$, $|s| \leq M$. Besides,

$$K_2 e^{-\alpha_2|\xi|} \leq K_2 e^{-\alpha_2(M^{0.5}+1)^{-1}(|s|^{0.5}+1)|\xi|}, \quad \text{if } |s| \leq M.$$

Therefore, we find that (3.4) has an exponential dichotomy in $E^{(k+0.5)\times k}(s)$ for $\mathcal{S} \cap \{|s| \le M\}$ and $\xi \in \mathbb{R}$.

By choosing larger $K$ and smaller $\alpha$, the result of (1) follows.

Observe that for any fixed $\sigma_0$, if $M$ is sufficiently large, we have $\{\text{Re}(s) \ge \sigma_0\} \subset \mathfrak{S}(M)$. The result of (2) then follows from that of (1). □

**Remark** Corollary 3.5 is valid only the exponential dichotomies are considered for $\xi \in \mathbb{R}$. Since it uses the "uniqueness of exponential dichotomy for $\xi \in \mathbb{R}$". Caution must be exercised when using it to prove, for example, Lemma 4.1 later.

**Lemma 3.6.** *Assume that $A$ is a constant matrix and $V$ is a constant scalar. Suppose that there exists $\sigma_0 > 0$ such that $Re\sigma(A) \le -\sigma_0$. Then for any $0 < \delta < \sigma_0$ and $k \ge 0$, (3.4) has an exponential dichotomy in $E^{(k+0.5)\times k}(s)$ for $Re(s) \ge -\sigma_0 + \delta$ and $\xi \in \mathbb{R}$.*

*Proof.* Let $\lambda$ be an eigenvalue for $J(s) = \begin{pmatrix} 0 & I \\ sI - A & -VI \end{pmatrix}$. Let $x + iy$ be an eigenvalue for $sI - A$. Then $\lambda = (-V \pm \sqrt{V^2 + 4x + 4iy})/2$. Using the fact $\text{Re}(s) \ge -\sigma_0 + \delta$, we find that counting the multiplicity, there are $n$ eigenvalues with

$$\text{Re}(\lambda) \le (-V - \sqrt{V^2 + 4\delta})/2 < 0,$$

and $n$ eigenvalues with

$$\text{Re}(\lambda) \ge (-V + \sqrt{V^2 + 4\delta})/2 > 0.$$

Therefore, (3.4) has an exponential dichotomy in $\mathbb{R}^{2n}$ for $\text{Re}(s) \ge -\sigma_0 + \delta$ and $\xi \in \mathbb{R}$. From Corollary 3.5, it has an exponential dichotomy in $E^{(k+0.5)\times k}(s)$ for the same $s$ and $\xi$. □

Consider

$$D_\xi \begin{pmatrix} \hat{u} \\ \hat{v} \end{pmatrix} = \begin{pmatrix} 0 & I \\ sI - A(\epsilon\xi) & -V(\epsilon\xi)I \end{pmatrix} \begin{pmatrix} \hat{u} \\ \hat{v} \end{pmatrix} + \begin{pmatrix} 0 \\ \hat{f} \end{pmatrix}. \tag{3.7}$$

When $\epsilon > 0$ is small, (3.7) is a system with slowly varying coefficients.

**Lemma 3.7.** *(Existence of exponential dichotomy in regular layers) Assume that the matrix $A(\xi)$ and the scalar $V(\xi)$ are $C^1$ bounded functions for $\xi \in \mathbb{R}$. Assume that*

$$|A|_{C^1} + |V|_{C^1} \le a,$$

*and there exist $\sigma_0 > 0$ such that for all $\xi \in \mathbb{R}$,*

$$Re\{\sigma(A(\xi))\} \le -\sigma_0 < 0.$$

Let $0 < \delta < \sigma_0$. Then there exists $\epsilon_0 > 0$ such that for $0 < \epsilon < \epsilon_0$, there (3.7) has an exponential dichotomy in $E^{(k+0.5) \times k}(s)$, $k \geq 0$, for $Re(s) \geq -\sigma_0 + \delta$ and $\xi \in \mathbb{R}$.

*Proof.* According to Corollary 3.5, (2), we only need to show that (3.7) has an exponential dichotomy in $\mathbb{R}^{2n}$ for any $s \in \{\text{Re}(s) \geq -\sigma_0 + \delta\}$ and $\xi \in \mathbb{R}$.

Fixed such $(s, \xi)$ as a parameter, from Lemma 3.6, system (3.7) has an exponential dichotomy in $\mathbb{R}^{2n}$ as a system with constant coefficients. Because of the bound on $|A| + |V|$, the constants $K, \alpha$ can be chosen independent of $\xi$, see [6]. We then use Proposition 1 from [6], page 50. Since

$$|D_\xi A(\epsilon\xi)| + |D_\xi V(\epsilon\xi)| \leq C\epsilon,$$

if $\epsilon_0 > 0$ is sufficiently small and $0 < \epsilon < \epsilon_0$, all the conditions in that proposition are satisfied. Therefore, (3.7) has an exponential dichotomy in $\mathbb{R}^{2n}$ for $s \in \{\text{Re}(s) \geq -\sigma_0 + \delta\}$ and $\xi \in \mathbb{R}$. □

**Lemma 3.8.** *Consider a closed liner operator $\mathcal{A} : L^2(\mathbb{R}) \to L^2(\mathbb{R})$ defined by*

$$\mathcal{A}u = u_{\xi\xi} + V(\xi)u_\xi + A(\xi)u.$$

*Assume that (3.4) has an exponential dichotomy in $E^{0.5 \times 0}(s)$ for $Re(s) \geq -\sigma_0$ and $\xi \in \mathbb{R}$, then $\mathcal{A}$ is a sectorial operator with*

$$|(\lambda - \mathcal{A})^{-1}| \leq \frac{C}{1 + |\lambda|},$$

*for all $Re\lambda \geq -\sigma_0$.*

*Proof.* For $g \in L^2(\mathbb{R})$, $\text{Re}\lambda \geq -\sigma_0$, we want to solve

$$\mathcal{A}u - \lambda u = g.$$

Convert it into a first order system,

$$\begin{pmatrix} u \\ v \end{pmatrix}_\xi = \begin{pmatrix} 0 & I \\ \lambda I - A(\xi) & -V(\xi)I \end{pmatrix} \begin{pmatrix} u \\ v \end{pmatrix} + \begin{pmatrix} 0 \\ g \end{pmatrix}.$$

From the existence of exponential dichotomy, the only $L^2$ solution can be written as

$$\begin{pmatrix} u \\ v \end{pmatrix} = \int_{-\infty}^\xi T(\xi, \zeta; \lambda) P_s(\zeta, \lambda) \begin{pmatrix} 0 \\ g(\zeta) \end{pmatrix} d\zeta + \int_\infty^\xi T(\xi, \zeta; \lambda) P_u(\zeta, \lambda) \begin{pmatrix} 0 \\ g(\zeta) \end{pmatrix} d\zeta.$$

Since $\begin{pmatrix} 0 \\ g \end{pmatrix} \in E^{0.5 \times 0}(\lambda)$, we have

$$|T(\xi, \zeta; \lambda) P_s(\zeta, \lambda) \begin{pmatrix} 0 \\ g(\zeta) \end{pmatrix}|_{E^{0.5 \times 0}(\lambda)} \leq K e^{-\alpha(|\lambda|^{0.5}+1)|\xi-\zeta|} | \begin{pmatrix} 0 \\ g(\zeta) \end{pmatrix} |_{E^{0.5 \times 0}(\lambda)}$$

$$\leq K e^{-\alpha(|\lambda|^{0.5}+1)|\xi-\zeta|} |g(\zeta)|_{\mathbb{R}^n}, \quad \xi \geq \zeta.$$

Similarly,

$$|T(\xi, \zeta; \lambda) P_u(\zeta, \lambda) \begin{pmatrix} 0 \\ g(\zeta) \end{pmatrix}|_{E^{0.5 \times 0}(\lambda)} \leq K e^{-\alpha(|\lambda|^{0.5}+1)|\xi-\zeta|} |g(\zeta)|_{\mathbb{R}^n}, \quad \xi \leq \zeta.$$

We now have

$$|u(\xi)|_{\mathbb{R}^n} \leq | \begin{pmatrix} u \\ v \end{pmatrix} |_{E^{0.5 \times 0}(\lambda)} (1 + |\lambda|^{0.5})^{-1}$$

$$\leq (1 + |\lambda|^{0.5})^{-1} (\int_{-\infty}^{\xi} + \int_{\xi}^{\infty}) K e^{-\alpha(1+|\lambda|^{0.5})|\xi-\zeta|} |g(\zeta)| d\zeta$$

$$\leq K(1 + |\lambda|^{0.5})^{-1} [e^{-\alpha(1+|\lambda|^{0.5})|\xi|} * |g(\xi)|].$$

Here "$*$" denotes the convolution in $\xi$. From a standard estimate on convolution,

$$|u|_{L^2} \leq K(1 + |\lambda|^{0.5})^{-1} |e^{-\alpha(1+|\lambda|^{0.5})|\xi|}|_{L^1} |g(\xi)|_{L^2}$$

$$\leq 2K(1 + |\lambda|^{0.5})^{-2} \alpha^{-1} |g|_{L^2}$$

$$\leq 2K(1 + |\lambda|)^{-1} \alpha^{-1} |g|_{L^2}.$$

The last estimate indicates that $\mathcal{A}$ is sectorial in $L^2(\mathbb{R})$. □

**Lemma 3.9.** *Let $\mathcal{S} \subset \mathbb{C}$ be a closed subset and $\mathcal{S} \cap \{|s| \geq M\} \subset \mathfrak{S}(M)$ for some $M > 0$. Assume that (3.4) has an exponential dichotomy in $\mathbb{R}^{2n}$ for $s \in \mathcal{S}$ and $\xi \in \mathbb{R}^-$ or $\mathbb{R}^+$ respectively. Let an operator $\mathcal{A}$ be defined as in Lemma 3.8. Assume that $\mathcal{S} \subset \rho(\mathcal{A})$. Then (3.4) has an exponential dichotomy in $E^{(k+0.5) \times k}(s)$ for $s \in \mathcal{S}$ and $\xi \in \mathbb{R}$.*

*Proof.* From Corollary 3.5, one only has to prove that (3.4) has an exponential dichotomy in $\mathbb{R}^{2n}$. Since the exponential dichotomies exist in $\mathbb{R}^-$ and $\mathbb{R}^+$ respectively, one only needs to show the transverse intersection of $\mathcal{R}P_s(0^+, s)$ and $\mathcal{R}P_u(0^-, s)$. If for some $s_0 \in \mathcal{S}$, this is not true. Then there exists $\phi \neq 0$ such that $\phi \in \mathcal{R}P_s(0^+, s_0) \cap \mathcal{R}P_u(0^-, s_0)$. Let $\begin{pmatrix} \hat{u} \\ \hat{v} \end{pmatrix} = T(\xi, 0; s_0)\phi$. Then $u = \mathcal{L}^{-1}\hat{u} \in L^2(\mathbb{R})$ and satisfies $s_0 u = \mathcal{A}u$. This is a contradiction to $s_0 \in \rho(\mathcal{A})$. □

**Lemma 3.10.** *Assume that $\gamma \in \mathbb{R}$, and (3.4) has an exponential dichotomy in $\mathbb{R}^{2n}$ for $Re(s) \geq \gamma$ and $\xi \in \mathbb{R}^-$ or $\mathbb{R}^+$ respectively. Assume* that $\mathcal{A}$ (see Lemma 3.8) has a simple eigenvalue $\lambda_0$, $Re(\lambda_0) > \gamma$, and no other eigenvalue in $Re(s) \geq \gamma$. Then (3.4) has an exponential dichotomy in $E^{(k+0.5)\times k}(s)$ for $\{Re(s) \geq \gamma\} \cap \{|s - \lambda_0| \geq \delta\}$ and $\xi \in \mathbb{R}$. Here $\delta$ is any positive constant. Let the constants of the exponential dichotomy be $K_\delta$, $\alpha_\delta$. Then $\alpha_\delta$ can be independent of $\delta$, but $K_\delta \to \infty$ as $\delta \to 0$. More precisely,

$$\mathcal{R}P_s(0^+, s) \oplus \mathcal{R}P_u(0^-, s) = \mathbb{R}^{2n}, \tag{3.8}$$

for all $\{Re(s) \geq \gamma$ and $s \neq \lambda_0\}$. The projections $P_1 + P_2 = I$, defined by the above splitting have a pole of order one at $s = \lambda_0$, i.e.,

$$|P_j| \leq C(1 + |s - \lambda_0|^{-1}), \quad j = 1, 2 \text{ for } Re(s) \geq \gamma.$$

*Proof.* We first show that $\{Re(s) \geq \gamma$ and $s \neq \lambda_0\} \subset \rho(\mathcal{A})$. Let $s$ be a point from the left hand side. Then (3.8) is valid, or $s \in \sigma(\mathcal{A})$, which is not true since $s \neq \lambda_0$. From (3.8), (3.4) has an exponential dichotomy for $\xi \in \mathbb{R}$. Using the same integral formula as in Lemma 3.7, we find that for each $g \in L^2(\mathbb{R})$, there exists a $u \in H^2(\mathbb{R})$ such that $\mathcal{A}u - su = g$. Thus, $s \in \rho(\mathcal{A})$.

The closed subset $\{Re(s) \geq \gamma \cap |s - \lambda_0| \geq \delta\} \subset \rho(\mathcal{A})$. Thus from Lemma 3.9, (3.4) has an exponential dichotomy in $E^{(k+0.5)\times k}(s)$ for $s \in \{Re(s) \geq \gamma \cap |s - \lambda_0| \geq \delta\}$ and $\xi \in \mathbb{R}$.

Consider $|s - \lambda_0| \leq \delta$, $\delta$ small. Based on a lemma due to Gardner & Jones [19], There exist $n$ independent vectors $\{w_i^-(s)\}_{i=1}^n$ that form a basis for $\mathcal{R}P_u(0^-, s)$ and $n$ independent vectors $\{w_i^+(s)\}_{i=1}^n$ that form a basis for $\mathcal{R}P_s(0^+, s)$. The vectors are analytic in $s$, so is

$$D(s) \stackrel{\text{def}}{=} \det(w_1^-(s) \ldots w_n^-(s) \, w_1^+(s) \ldots w_n^+(s)).$$

The results from [19] also assert that

(1) $D(s) = 0$ in $|s - \lambda_0| \leq \delta$ if and only if $s \in \sigma_p(\mathcal{A})$;

(2) The order of the roots of $D$ at $\lambda = \lambda_0$ equals the algebraic multiplicity of $\lambda_0$ as an eigenvalue of $\mathcal{A}$.

We infer that $s = \lambda_0$ is a simple root of $D$ since $\lambda_0$ is a simple eigenvalue.

For $w \in \mathbb{R}^{2n}$, $|w| = 1$, consider

$$w = (w_1^-(s) \ldots w_n^-(s) \, w_1^+(s) \ldots w_n^+(s)) \begin{pmatrix} c_1(s) \\ \vdots \\ c_{2n}(s) \end{pmatrix}.$$

From Cramer's rule, we easily find that $c_j(s) = O(|D(s)|^{-1}) = O(|s - \lambda_0|^{-1})$. Therefore, $P_1 w = \sum_{i=1}^n w_i^-(s) c_i(s) = O(|s - \lambda_0|^{-1})$. Similarly $P_2 w = O(|s - \lambda_0|^{-1})$.

The desired estimates on $P_1$ and $P_2$ follow by combine estimates in $\{\operatorname{Re}(s) \geq \gamma \cap |s - \lambda_0| \geq \delta\}$ and $\{|s - \lambda_0| \leq \delta\}$. $\qquad \square$

**Lemma 3.11.** *Assume that $\mathcal{A}$ is the operator defined in Lemma 3.8 and is sectorial in $L^2(\mathbb{R})$ with constants $\alpha_0 > 0$, $C > 0$ such that*

$$\Sigma = \{|\arg(\lambda + \alpha_0)| \leq \theta, \ \pi/2 < \theta < \pi\} \backslash \{-\alpha_0\} \subset \rho(\mathcal{A}),$$
$$|(\lambda - \mathcal{A})^{-1}|_{L^2} \leq C_0 |\lambda + \alpha_0|^{-1}, \quad \text{for } \lambda \in \Sigma.$$

*Let $\gamma \leq 0$ with $|\gamma| \leq \alpha_0/4$. Let $g \in L^2(\mathbb{R} \times \mathbb{R}^+, \gamma)$ and $u_0 \in H^1(\mathbb{R})$. Consider the initial value problem,*

$$u_\tau = \mathcal{A}u + g,$$
$$u(\xi, 0) = u_0(\xi), \quad \xi \in \mathbb{R}, \ \tau \geq 0.$$

*Then the solution $u \in H^{2,1}(\mathbb{R} \times \mathbb{R}^+, \gamma)$, and*

$$|u|_{H^{2,1}(\gamma)} \leq C(|u_0|_{H^1} + |g|_{L^2(\gamma)}),$$

*where $C = \sqrt{2} C_0 (1 + \dfrac{1}{\alpha_0 + 3\gamma})$.*

*Proof.* $B = \mathcal{A} - \gamma$ is a sectorial operator with

$$|(\lambda - B)^{-1}|_{L^2} \leq \frac{C_0}{|\lambda + \alpha_0 + \gamma|},$$

for $\lambda \in \{|\arg(\lambda + \alpha_0 + \gamma)| \leq \theta, \ \pi/2 < \theta < \pi\} \backslash \{-\alpha_0 - \gamma\}$. Suppose now $\operatorname{Re}(\lambda) \geq \gamma$,

$$\begin{aligned}
|\lambda + \alpha_0 + \gamma|^{-1} &\leq [(\lambda - \gamma)^2 + (\alpha_0 + 2\gamma)^2]^{-1/2} \\
&\leq \sqrt{2}[|\lambda - \gamma| + (\alpha_0 + 2\gamma)]^{-1} \\
&\leq \sqrt{2}[|\lambda| + \alpha_0 + 3\gamma]^{-1} \\
&\leq (1 + (\alpha_0 + 3\gamma)^{-1}) \frac{\sqrt{2}}{|\lambda| + 1}.
\end{aligned}$$

Therefore

$$|(\lambda - B)^{-1}|_{L^2} \leq C \frac{1}{1 + |\lambda|}, \tag{3.9}$$

where $C$ is the constant defined in the lemma.

The solution of the initial value problem satisfies

$$e^{-\gamma t} u = e^{Bt} u_0 + e^{Bt} * G(t),$$

where "$*$" denotes the convolution in $t$ and $G(t) = e^{-\gamma t} g(t)$, with $|G|_{L^2} = |g|_{L^2(\gamma)}$. From (3.9), and [28]

$$|e^{Bt} * G(t)|_{H^{2,1}} \leq C|G|_{L^2},$$
$$|e^{Bt} u_0|_{H^{2,1}} \leq C|u_0|_{H^1}.$$

The desired result then follows easily. □

## 4. Coordinates Change and Linearization

We first construct new coordinates in the region

$$\Sigma^i \cap [\bar{t}, \bar{t} + \Delta t], \bar{t} \geq 0, \quad \text{or} \quad \Sigma^i \cap [t_f, \infty),$$

to straighten the boundaries $\Gamma^{i-1}$ & $\Gamma^i$. Here $\Delta t = \epsilon \Delta \tau$, $\Delta \tau > 0$ and $t_f > 0$. Let $0 \leq \theta(x) \leq 1$ be a $C^\infty$ function, $\theta(0) = 1/2$, $\theta(x) = 1$ for $x \geq 1$ and $\theta(x) = 0$ for $x \leq -1$. Let $m^i(t) = (y^{i-1}(t) + y^i(t))/2$ and $\Delta y^i(t) = y^i(t) - y^{i-1}(t)$. For each fixed $\bar{t} \geq 0$ or $\bar{t} = \infty$, let $I = [\bar{t}, \bar{t} + \Delta t]$ or $I = [t_f, \infty)$ respectively. Define a change of coordinates in $\Sigma^i \cap I$,

$$y = y^*(x, t, \bar{t}, i) = x - \Xi(x, t, \bar{t}, i),$$

$$\Xi(x, t, \bar{t}, i) = (y^i(t) - y^i(\bar{t}))\theta(\frac{x - m^i(t)}{\Delta y^i(t)/4}) + (y^{i-1}(t) - y^{i-1}(\bar{t}))(1 - \theta(\frac{x - m^i(t)}{\Delta y^i(t)/4})).$$

Note that $y^i(t)$ really depends on $\epsilon$, so is $y^*$ and $\Xi$. For typing convenience, we will also drop $\bar{t}, i$ in $y^*$, $\Xi$ if no confusion should arise.

From its definition, it is clear that $\Xi = 0$ for $t = \bar{t}$.

Let $C > 0$ be a given constant. For the case $\bar{t} = \infty$, from §2, **H1**, there exists $t_f$ sufficiently large such that for $I = [t_f, \infty)$,

$$|y^i(t) - y^i(\bar{t})| + |y^{i-1}(t) - y^{i-1}(\bar{t})| \leq C, \quad t \in I, \quad (4.1)$$

For the case $0 < \bar{t} < \infty$, if $\epsilon > 0$ is sufficiently small, then for $I = [\bar{t}, \bar{t} + \Delta t]$, we also have (4.1). If $C$ is sufficiently small, with the corresponding $t_f$ or $\Delta t$, we have

$$\partial_x \Xi(x, t, \bar{t}, i) \leq \frac{1}{2}, \quad (x, t) \in \Sigma^i \cap I.$$

Thus, $y = y^*(x, t, \bar{t}, i)$ has a smooth inverse $x = x^*(y, t, \bar{t}, i)$, $y^{i-1}(\bar{t}) \leq y \leq y^i(\bar{t})$, $t \in I$.

The change of coordinates has the following properties:

(1) $y = x$ at $t = \bar{t}$;

(2) A neighborhood of $\Gamma^i$ in $\Sigma^i \cap I$ is shifted by the amount $y^i(t) - y^i(\bar{t})$. Thus in the new coordinates, $\Gamma^i = \{y = y^i(\bar{t})\}$. For the same reason, $\Gamma^{i-1} = \{y = y^{i-1}(\bar{t})\}$, and $\Sigma^i \cap I = \{y^{i-1}(\bar{t}) \leq y \leq y^i(\bar{t})\}$. Also, if $|x - y^i(t)|$ or $|x - y^{i-1}(t)| \leq \Delta y^i(t)/4$, then

$$\partial_x^k \Xi(x, t, \bar{t}, i) = 0, \quad k \geq 1.$$

(3) The curve that bisects the region is straightened, $y^*(m^i(t), t) = m^i(\bar{t})$.

(4) (a) In a finite interval $I = [\bar{t}, \bar{t} + \Delta t]$, $|\partial_y^k \partial_t^h x^*(y, t)| + |\partial_x^k \partial_t^h \Xi(x, t)| \leq C_{kh}$ uniformly with respect to $\bar{t} \geq 0$.

(b) In the interval $I_\infty = [t_f, \infty)$, $|\partial_y^k \partial_t x^*(y,t)| + |\partial_x^k \partial_t \Xi(x,t)| \leq C_k e^{-\bar\gamma t_f}$.

The proof of (4) (a) is based on the continuity of $D^h y^i(t)$. The proof of (4) (b) is based on $|Dy^i(t)| \leq C e^{-\bar\gamma t}$, which in turn is based on **H1**.

We now choose $t_f$ larger, if necessary, so that there exists an integer $r > 0$ such that $t_f = r\Delta t$. Let $I_j = [j\Delta t, (j+1)\Delta t]$, $0 \leq j \leq r-1$ and $I_\infty = [t_f, \infty)$. A partition of the time axis is made by $I_j$, $0 \leq j \leq r-1$ and $I_\infty$. Define a change of coordinates by $y = y^*(x, t, \bar t, i)$ with $\bar t = j\Delta t$ in $\Sigma^i \cap I_j$ and $\bar t = \infty$ in $\Sigma^i \cap I_\infty$. Using the new variable, a solution of (1.1) in regular or singular regions satisfies

$$u(x,t) = U(y,t) = U(x - \Xi(x,t), t),$$
$$\epsilon(U_t - U_y \Xi_t) = \epsilon^2((1 - \Xi_x)^2 U_{yy} - \Xi_{xx} U_y) + f(U, x^*, \epsilon).$$

We now construct a new coordinate system in the region $\Sigma^i \cap I_j$, $0 \leq j \leq r-1$ or $j = \infty$. Let $\bar t = j\Delta t$ in $\Sigma^i \cap I_j$, including $j = \infty$. Let $L_j^i(\epsilon) = \Delta y^i(\bar t)/2\epsilon$ and $\Omega_j^i = (-L_j^i(\epsilon), L_j^i(\epsilon))$.

$$\mathfrak{R}(0) \to \mathfrak{R}^i(1), \, : \Sigma^i \cap I_j \ni (x,t) \to (\xi, \tau) \in \Omega_j^i \times I_\tau^j,$$
$$y = x - \Xi(x, t, \bar t, i),$$
$$\xi = [y - m^i(\bar t)]/\epsilon,$$
$$\tau = \begin{cases} (t - j\Delta t)/\epsilon, & j \neq \infty, \\ (t - t_f)/\epsilon, & j = \infty, \end{cases}$$
$$I_\tau^j = \begin{cases} [0, \Delta\tau], & \text{in } \Sigma^i \cap I_j, \, 0 \leq j \leq r-1, \\ [0, \infty) & \text{in } \Sigma^i \cap I_\infty. \end{cases}$$

Using the coordinates $\mathfrak{R}^i(1)$, we denote the solution in $\Omega_j^i \times I_\tau^j$, $0 \leq j \leq r-1$, or $\infty$ by $U^i(\xi, \tau)$. The equation for $U^i$ is

$$U_\tau^i = U_{\xi\xi}^i + \Xi_t(x^*(m^i(\bar t) + \epsilon\xi, \bar t), \bar t)U_\xi^i + f(U^i, x^*(m^i(\bar t) + \epsilon\xi, t), \epsilon) \tag{4.2}$$
$$+ \mathcal{N}_1^i(U^i, \xi, t).$$

Here $\mathcal{N}_1^i = 0$ if $t = \bar t$. $\mathcal{N}_1^i$ is linear in $U^i$:

$$\mathcal{N}_1^i(U^i, \xi, t) = [-2\Xi_x(x^*, t) + \Xi_x^2(x^*, t)]U_{\xi\xi}^i - \epsilon\Xi_{xx}(x^*, t)U_\xi^i$$
$$+ [\Xi_t(x^*(m^i(\bar t) + \epsilon\xi, t), t) - \Xi_t(x^*(m^i(\bar t) + \epsilon\xi, \bar t), \bar t)]U_\xi^i.$$

Here $x^*$ denotes $x^*(m^i(\bar t) + \epsilon\xi, t)$ and $t = \bar t + \epsilon\tau$ if $\bar t = j\Delta t$ or $t = t_f + \epsilon\tau$ if $\bar t = \infty$.

Using (4), (a), it is not hard to verify that if $\bar t = j\Delta t$, $0 \leq j \leq r-1$, then

$$|\mathcal{N}_1^i|_{L^2(\Omega_j^i \times I_\tau^j)} \leq C\Delta t |U^i|_{H^{2,1}(\Omega_j^i \times I_\tau^j)}.$$

Using (4), (b), if $\bar{t} = \infty$, $j = \infty$, then for any $\gamma < 0$,

$$|\mathcal{N}_1^i|_{X(\Omega_\infty^i \times I_\tau^j, \gamma)} \leq C e^{-\bar{\gamma} t_f} |U^i|_{X^{2,1}(\Omega_\infty^i \times I_\tau^j, \gamma)}.$$

Let $U^i = w^i + u^i$. Linearizing (4.2) around $w^i(m^i(\bar{t}) + \epsilon \xi, \bar{t}, \epsilon)$, where $w^i$ is the formal approximation, we have

$$u_\tau^i = u_{\xi\xi}^i + V^i(\xi) u_\xi^i + A^i(\xi) u^i + \mathcal{N}^i(u^i, \xi, t, \epsilon), \quad (\xi, \tau) \in \Omega_j^i \times I_\tau^j. \tag{4.3}$$

Here $A^i(\xi) = f_u(w^i(m^i(\bar{t}) + \epsilon \xi, \bar{t}, \epsilon), m^i(\bar{t}) + \epsilon \xi, \epsilon)$, $V^i(\xi) = \Xi_t(m^i(\bar{t}) + \epsilon \xi, \bar{t})$. One can verify that $V^{2\ell}(\xi) = \eta_t^\ell(\bar{t}, \epsilon)$ is the speed of the wave front, and is independent of $\xi$. $V^{2\ell+1}$ depends slowly on $\xi$. $V^i = 0$ if $\bar{t} = \infty$. Denote $w^i(t) = w^i(x^*(m^i(\bar{t}) + \epsilon \xi, t), t, \epsilon)$,

$$\mathcal{N}^i = \mathcal{N}_1^i + g^i + [f(w^i(t) + u^i, x^*(m^i(\bar{t}) + \epsilon \xi, t), \epsilon) - f(w^i(t), x^*(m^i(\bar{t}) + \epsilon \xi, t), \epsilon)$$
$$- f_u(w^i(t), x^*(m^i(\bar{t}) + \epsilon \xi, t), \epsilon) u^i]$$
$$+ [f_u(w^i(t), x^*(m^i(\bar{t}) + \epsilon \xi, t), \epsilon) - f_u(w^i(\bar{t}), m^i(\bar{t}) + \epsilon \xi, \bar{t}, \epsilon)] u^i.$$

When $0 \leq j \leq r - 1$, we have, in $\Omega_j^i \times I_\tau^j$,

$$|\mathcal{N}^i|_{L^2} \leq |g^i|_{L^2} + C(\Delta t |u^i|_{H^{2,1}} + |u^i|_{H^{2,1}}^2),$$
$$|\mathcal{N}^i(\bar{u}^i) - \mathcal{N}^i(\bar{\bar{u}}^i)|_{L^2} \leq C(\Delta t + |\bar{u}^i|_{H^{2,1}} + |\bar{\bar{u}}^i|_{H^{2,1}})|\bar{u}^i - \bar{\bar{u}}^i|_{H^{2,1}}. \tag{4.4}$$

When $j = \infty$, we have, in $\Omega_\infty^i \times I_\tau^j$, $\gamma < 0$,

$$|\mathcal{N}^i|_{X(\gamma)} \leq |g^i|_{X(\gamma)} + C(e^{-\bar{\gamma} t_f} |u^i|_{X^{2,1}(\gamma)} + |u^i|_{X^{2,1}(\gamma)}^2),$$
$$|\mathcal{N}^i(\bar{u}^i) - \mathcal{N}^i(\bar{\bar{u}})|_{X(\gamma)} \leq C(e^{-\bar{\gamma} t_f} + |\bar{u}^i|_{X^{2,1}(\gamma)} + |\bar{\bar{u}}^i|_{X^{2,1}(\gamma)})|\bar{u}^i - \bar{\bar{u}}^i|_{X^{2,1}(\gamma)}. \tag{4.5}$$

We are led to solve system (4.3) with $\bar{t} = j\Delta t$, $I_\tau^j = [0, \Delta \tau]$ if $0 \leq j \leq r - 1$, or $I_\tau^j = \mathbb{R}^+$ if $j = \infty$. The nonlinear term $\mathcal{N}^i$ satisfies (4.4) or (4.5). The solution satisfies jump conditions (1.8) in the form

$$\binom{I}{D_\xi} [u^i(L_j^i(\epsilon), \tau, \epsilon) - u^{i+1}(-L_j^{i+1}(\epsilon), \tau, \epsilon)] = \delta^i(\tau), \tag{4.6}$$

initial conditions

$$u^i(\xi, 0) = u_0^i(\xi), \tag{4.7}$$

and compatibility conditions

$$\pi_1 \delta^i(0) = u_0^i(L_j^i(\epsilon), 0) - u_0^{i+1}(-L_j^{i+1}(\epsilon), 0).$$

Here $\pi_1 : \mathbb{R}^{2n} \to \mathbb{R}^n$, $(u, v)^\tau \to u$, for $u, v \in \mathbb{R}^n$, is a projection. For $\delta^i = (\delta_1^i, \delta_2^i)^\tau \in H^{0.75 \times 0.25}(I_\tau^j)$, observe only the trace of $\delta_1^i$ is defined at $\tau = 0$. Therefore, the compatibility condition is posed only on $\pi_1 \delta^i = \delta_1^i$.

Let us discuss properties of (4.3), in regular or internal layers, and in finite or infinite time intervals. Assume the the coefficients of (4.3) have been extended continuously from $\xi \in \Omega_j^i$ to $\xi \in \mathbb{R}$ by constants. This is equivalent to introducing $\tilde{\xi}$ in internal layers, see §2. If, after freezing the coefficients $A^i(\xi)$ and $V^i(\xi)$ at $\xi = \xi_0$, (4.8') below has an exponential dichotomy for $\xi \in \mathbb{R}$, then the projections for the dichotomy are denoted by $\bar{P}_s^i(\xi_0)$ and $\bar{P}_u^i(\xi_0)$.

Let $\mathcal{A}^i u = u_{\xi\xi} + V^i(\xi) u_\xi + A^i(\xi) u$. The dual system associate to $u_\tau = \mathcal{A}^i u$ is (4.8'). When $i = 2\ell - 1$, for each fixed $\xi$, $A^i(\xi)$ satisfies **H2**, and $|\partial_\xi A^i(\xi)| + |\partial_\xi V^i(\xi)| = O(\epsilon)$. From Lemma 3.7, we have

**P 1.** In the regular layers, (4.8') has an exponential dichotomy in $E^{(k+0.5)\times k}(s)$ for $\mathrm{Re}(s) \geq -\bar{\sigma} + \sigma_1$, $\sigma_1 > 0$, and $\xi \in \mathbb{R}$. Also the projections of the dichotomy at any $\xi_0 \in \mathbb{R}$ are close to those obtained by freezing the coefficients, i.e., $P_s^i(\xi_0, s) = \bar{P}_s^i(\xi_0, s) + O(\epsilon)$, $P_u^i(\xi_0, s) = \bar{P}_u^i(\xi_0, s) + O(\epsilon)$.

Based on Lemma 3.8, we have

**P 2.** In the regular layer, $|(\lambda - \mathcal{A}^i)^{-1}|_{L^2} \leq C/(1 + |\lambda|)$, for $\mathrm{Re}(\lambda) \geq -\bar{\sigma} + \sigma_1$.

For $i = 2\ell$, if the coefficients are constants, $A^{i+1}(-L_j^{i+1}(\epsilon))$ and $V^{i+1}(-L_j^{i+1}(\epsilon))$, system (4.8') has an exponential dichotomy for $\xi \in \mathbb{R}$, due to **H2** and Lemma 3.6. Similar assertions holds for (4.8') with coefficients $A^{i-1}(L_j^{i-1}(\epsilon))$ and $V^{i-1}(L_j^{i-1}(\epsilon))$. Using Lemma 3.3, and **H3**, we can show

**P 3.** In the internal layers, there exists a large constant $N > 0$ such that the dual system (4.8') has an exponential dichotomy in $E^{(k+0.5)\times k}(s)$ for $\mathrm{Re}(s) \geq -\bar{\sigma} + \sigma_1$, $\sigma_1 > 0$, and $\xi \in [-\infty, -N]$ or $[N, \infty)$ respectively. The projections are close to $\bar{P}_s^{i-1}(L_j^{i-1}(\epsilon), s)$ and $\bar{P}_u^{i-1}(L_j^{i-1}(\epsilon), s)$ for $\xi \leq -N$. They are close to $\bar{P}_s^{i+1}(-L_j^{i+1}(\epsilon), s)$ and $\bar{P}_u^{i+1}(-L_j^{i+1}(\epsilon), s)$ for $\xi \geq N$.

Based on **P3**, (4.8') has exponential dichotomies in $\mathbb{R}^{2n}$ for $\xi \in \mathbb{R}^-$ and $\mathbb{R}^+$ respectively. They are natural extensions form the dichotomies in $(-\infty, -N]$ and $[N, \infty)$. From **H4** and Lemma 3.10, we have

**P 4.** In the internal layers, (4.8') has an exponential dichotomy in $E^{(k+0.5)\times k}(s)$ for $s \in \{\mathrm{Re}(s) \geq -\bar{\sigma} + \delta \cap \{|s - \lambda^i(\epsilon)| \geq \delta\}\}$ for all $\delta > 0$ and $\xi \in \mathbb{R}$. The projections defined by the splitting

$$\mathcal{R}P_s^i(0^+, s) \oplus \mathcal{R}P_u^i(0^-, s) = \mathbb{R}^{2n}$$

are bounded by $C(1 + \dfrac{1}{|s - \lambda^i(\epsilon)|})$. Further more $\lambda^i(\epsilon) = \epsilon\lambda_0^i(\bar{t}) + O(\epsilon^2)$.
In the limiting case, we also have $\lambda_0^i(\infty) \leq \bar{\lambda} < 0$, $i \in \mathbb{Z}$.

**Remark** Exponential dichotomies are not unique in semi-infinite intervals $[N, \infty)$ and $(-\infty, -N]$. In particular, the projections that define the exponential dichotomies in **P3** and **P4** are different. However, the unstable subspace at $-N$, $\mathcal{R}P_u^i(-N, s)$, and the stable subspace at $N$, $\mathcal{R}P_s^i(N, s)$, are unique.

We now present an important lemma that is used in §5 and §6.

Consider a sequence of equations, $i \in \mathbb{Z}$,

$$u_\tau^i = u_{\xi\xi}^i + V^i(\xi)u_\xi^i + A^i(\xi)u^i, \quad \xi \in \Omega_j^i, \tau \in I_\tau^j \tag{4.8}$$

$$u^i(\xi, 0) = 0, \tag{4.9}$$

$$\begin{pmatrix} I \\ D_\xi \end{pmatrix}(u^i(L_j^i(\epsilon), \cdot) - u^{i+1}(-L_j^{i+1}(\epsilon), \cdot) = \delta^i \in H_0^{0.75 \times 0.25}(\gamma). \tag{4.10}$$

The compatibility $\pi_1\delta^i(0) = 0$ is clearly satisfied.

The dual systems for (4.8)–(4.10) are

$$D_\xi \begin{pmatrix} \hat{u}^i \\ \hat{v}^i \end{pmatrix} = \begin{pmatrix} 0 & I \\ sI - A^i(\xi) & -V^i(\xi)I \end{pmatrix} \begin{pmatrix} \hat{u}^i \\ \hat{v}^i \end{pmatrix} \tag{4.8'}$$

$$\begin{pmatrix} I \\ D_\xi \end{pmatrix}(\hat{u}^i(L_j^i(\epsilon), s) - \hat{u}^{i+1}(-L_j^{i+1}(\epsilon), s)) = \hat{\delta}^i(s), \tag{4.10'}$$

where $\hat{\delta}^i \in \mathcal{H}^{0.75 \times 0.25}(\gamma)$. Let

$$B = \sup_{i,\xi}\{|A^i(\xi)| + |V^i(\xi)|\}.$$

We first derive a Gronwall type inequality that governs the growth of solutions of (4.8'). A solution $\phi(\xi, s)$ of (4.8') satisfies the integral equation,

$$\phi(\xi, s) = e^{J(s)\xi}\phi(0, s) + \int_0^\xi e^{J(s)(\xi-\zeta)} \begin{pmatrix} 0 & 0 \\ -A^i(\zeta) & -V^i(\zeta)I \end{pmatrix} \phi(\zeta, s)d\zeta.$$

Consider $s \in \mathfrak{S}(M)$, $M = 1$ first. See (3.5) for the definition of $\mathfrak{S}(M)$. It is known that $J(s) = \begin{pmatrix} 0 & I \\ sI & 0 \end{pmatrix}$ has eigenvalues $\lambda = \pm\sqrt{s}$, each has an $n$-dimensional eigenspace. The spectral projections to the eigenspaces are bounded uniformly with respect to $s$ in the norm of $E^{0.75 \times 0.25}(s)$. Thus for $\xi \geq 0$, we have,

$$|\phi(\xi, s)| \leq Ce^{\sqrt{|s|}\xi}|\phi(0, s)| + C\int_0^\xi e^{\sqrt{|s|}(\xi-\zeta)}B|\phi(\zeta, s)|d\zeta,$$

where the norms are in $E^{0.75 \times 0.25}(s)$. The constants $C$ are related to the norms of the spectral projections. From Gronwall's inequality,

$$|\phi(\xi, s)| \leq C_1 e^{(\sqrt{|s|} + D_1)\xi}|\phi(0, s)|, \quad \xi \geq 0, \ D_1 = CB. \tag{4.11}$$

Now consider $s \in \{\mathrm{Re}(s) \geq -\bar{\sigma}\} \backslash \mathfrak{S}(M)$, a compact set in $\mathbb{C}$. By Gronwall again, we have, in the $\mathbb{R}^{2n}$ norm,

$$|\phi(\xi, s)| \leq C_2 e^{D_2 \xi}|\phi(0, s)|, \quad \xi \geq 0. \tag{4.12}$$

The constants $C_1$, $C_2$, $D_1$, $D_2$ are all positive. Results similar to (4.11) and (4.12) can also be derived for $\xi < 0$. By combining (4.11) and (4.12), we have,

$$|\phi(\xi, s)|_{E^{0.75 \times 0.25}(s)} \leq C e^{\eta(\sqrt{|s|}+1)|\xi|}|\phi(0, s)|_{E^{0.75 \times 0.25}(s)}, \tag{4.13}$$

for $\xi \in \mathbb{R}$ and $\mathrm{Re}(s) \geq -\bar{\sigma}$. Here the constants $C$ and $\eta$ depend on $\bar{\sigma}$ and $B$ only.

**Lemma 4.1.** *Assume that (4.8) satisfies* **P1** *and* **P2** *in regular layers* ($i = 2\ell - 1$), *and satisfies* **P3** *and* **P4** *in internal layers* ($i = 2\ell$). *Assume either one of the following conditions is valid.*

*(1)* $j = \infty$, $I_\tau^j = \mathbb{R}^+$, $\gamma = \gamma_0 \epsilon$, $\gamma_0 < 0$ *and* $\gamma - \sup_i\{|\lambda^i(\epsilon)|\} \geq O_1 \epsilon$ *for some* $O_1 > 0$.

*(2)* $0 \leq j \leq r - 1$, $I_\tau^j = [0, \Delta\tau]$, $\gamma$ *is a constant, independent of* $\epsilon$, *and* $-\bar{\sigma} < \gamma \leq 0$.

*Then, in any of the two cases, there exists* $\epsilon_0 > 0$ *such that for* $0 < \epsilon < \epsilon_0$, *(4.8)–(4.10) has a unique solution* $\{u^i\}_{-\infty}^{\infty}$ *with* $u^i \in H^{2,1}(\Omega_j^i \times I_\tau^j, \gamma)$. *Moreover,*

$$|u^i|_{H^{2,1}(\Omega_j^i \times I_\tau^j, \gamma)} \leq C|\{\delta^i\}|_{H_0^{0.75 \times 0.25}(I_\tau^j, \gamma)}. \tag{4.14}$$

*Proof.* Consider case (1) first. Let $-\sigma_0 = -\bar{\sigma} + \sigma_1 < \inf\{\lambda^i(\epsilon)\}$. Then $-\sigma_0 < \gamma$. By **P1**, (4.8') has exponential dichotomies in regular layers for $\mathrm{Re}(s) \geq -\sigma_0$ and $\xi \in \mathbb{R}$.

By **P3**, in internal layers, there exists exponential dichotomies for $\mathrm{Re}(s) \geq -\sigma_0$ and $\xi \in (-\infty, -N)$ or $[N, \infty)$. The range of the projections at the boundary $\Gamma^i$, $\mathcal{R}P_s(-L_j^{i+1}(\epsilon), s)$ and $\mathcal{R}P_u^i(L_j^i(\epsilon), s)$, are close to the stable and unstable eigenspaces of a constant system. Thus, the projections defined by the following splitting

$$\mathcal{R}P_s^{i+1}(-L_j^{i+1}(\epsilon), s) \oplus \mathcal{R}P_u^i(L_j^i(\epsilon), s),$$

are uniformly bounded with respect to $\epsilon$. For $\hat{\delta}^i \in \mathcal{H}^{0.75 \times 0.25}(\gamma)$, there exist unique $\phi_u^i(L_j^i(\epsilon), s) \in \mathcal{R}P_u^i(L_j^i(\epsilon), s)$ and $\phi_s^{i+1}(-L_j^{i+1}(\epsilon), s) \in$

$\mathcal{R}P_s^{i+1}(-L_j^{i+1}(\epsilon), s)$, with

$$\phi_u^i(L_j^i(\epsilon), s) - \phi_s^{i+1}(-L_j^{i+1}(\epsilon), s) = \hat{\delta}^i(s).$$

$$|\phi_u^i(L_j^i(\epsilon), \cdot)|_{\mathcal{H}^{0.75 \times 0.25}(\gamma)} + |\phi_s^{i+1}(-L_j^{i+1}(\epsilon), \cdot)|_{\mathcal{H}^{0.75 \times 0.25}(\gamma)} \leq C|\hat{\delta}^i|_{\mathcal{H}^{0.75 \times 0.25}(\gamma)}.$$

In a regular layer, let

$$\phi^i = T^i(\xi, -L_j^i(\epsilon), s)\phi_s^i(-L_j^i(\epsilon), s) + T^i(\xi, L_j^i(\epsilon), s)\phi_u^i(L_j^i(\epsilon), s), \quad |\xi| \leq L_j^i(\epsilon).$$

In an internal layer, first consider

$$\phi = T^i(\xi, -L_j^i(\epsilon), s)\phi_s^i(-L_j^i(\epsilon), s), \quad \xi \in (-L_j^i(\epsilon), 0].$$

Since (4.8') has an exponential dichotomy for $\xi \in (-\infty, -N]$ or $[N, \infty)$, therefore $u = \mathcal{L}^{-1}(\pi_1 \phi)$, $\xi \in (-L_j^i(\epsilon), -N)$ is a solution of (4.8) with $u \in H^{2,1}([-L_j^i(\epsilon), -N] \times I_\tau^j, \gamma)$, according to Lemma 3.1. Moreover,

$$|\phi(-N, s)|_{E^{0.75 \times 0.25}(s)} \leq Ke^{-\alpha(\sqrt{|s|}+1)(L_j^i(\epsilon)-N)}|\phi_s^i(-L_j^i(\epsilon), s)|_{E^{0.75 \times 0.25}(s)}.$$

We want to show $u \in H^{2,1}([-N, 0] \times I_\tau^j, \gamma)$ also. The exponential dichotomy used above has not been extended to $[-N, 0]$. However, using inequality (4.13), we can show

$$|\phi(\xi, s)|_{E^{0.75 \times 0.25}(s)} \leq Ce^{\eta(\sqrt{|s|}+1)(\xi+N)}|\phi(-N, s)|_{E^{0.75 \times 0.25}(s)}, \quad -N \leq \xi \leq 0.$$

Let $s = \sigma + i\omega$ with $\sigma > \gamma$. Let $\epsilon$ be sufficiently small such that $\alpha(L_j^i(\epsilon) - N) \geq \eta N$. We have, as a function of $\omega$ and $\xi$,

$$\begin{aligned}
|\hat{u}|_{L^2}^2 + |s\hat{u}|_{L^2}^2 &\leq \int_{-\infty}^{\infty}\int_{-N}^0 (|s|^2 + 1)|\hat{u}(\xi, s)|^2 d\xi d\omega \\
&\leq \int_{-\infty}^{\infty}\int_{-N}^0 (|s|^{0.5} + 1)|\phi(\xi, s)|_{E^{0.75 \times 0.25}(s)}^2 d\xi d\omega \\
&\leq \int_{-\infty}^{\infty}\int_{-N}^0 (|s|^{0.5} + 1)e^{2\eta(\sqrt{|s|}+1)(\xi+N)}|\phi(-N, s)|_{E^{0.75 \times 0.25}(s)}^2 d\xi d\omega \\
&\leq C\int_{-\infty}^{\infty} e^{2\eta(\sqrt{|s|}+1)N}K^2 e^{-2\alpha(\sqrt{|s|}+1)(L_j^i(\epsilon)-N)}|\phi_s^i(-L_j^i(\epsilon), s)|_{E^{0.75 \times 0.25}(s)}^2 d\omega \\
&\leq CC_0 \int_{-\infty}^{\infty} |\phi_s^i(-L_j^i(\epsilon), s)|_{E^{0.75 \times 0.25}(s)}^2 d\omega \\
&\leq CC_0 |\phi_s^i|_{\mathcal{H}^{0.72 \times 0.25}(\gamma)}^2.
\end{aligned}$$

Here $C_0 \leq e^{2\eta N - 2\alpha(L_j^i(\epsilon)-N)} \to 0$ as $L_j^i(\epsilon) \to \infty$.

Let $U^1(\xi, s) = T^i(\xi, -L_j^i(\epsilon), s)\phi_s^i(-L_j^i(\epsilon), s)$, $U^2(\xi, s) = T^i(\xi, L_j^i(\epsilon), s)\phi_u^i(L_j^i(\epsilon), s)$. We have shown that $\mathcal{L}^{-1}\pi_1 U^1 \in H^{2,1}([-L_j^i(\epsilon), 0] \times I_\tau^j, \gamma)$. Similarly, $\mathcal{L}^{-1}\pi_1 U^2 \in H^{2,1}([0, L_j^i(\epsilon)] \times I_\tau^j, \gamma)$. From **P4**, when $\text{Re}(s) > \gamma > \sup_i \lambda^i(\epsilon)$, there exists an exponential dichotomy for $\xi \in \mathbb{R}$. This is not the same exponential dichotomy guaranteed by **P3**, see the remark after **P4**. Denote the projections associated to that dichotomy by $\tilde{P}_s$ and $\tilde{P}_u$. Let

$$
\begin{aligned}
\phi_s^i(0^+, s) &= \tilde{P}_s^i(0, s)(U^1(0, s) - U^2(0, s)), \\
\phi_u^i(0^-, s) &= \tilde{P}_u^i(0, s)(U^1(0, s) - U^2(0, s)).
\end{aligned}
\tag{4.15}
$$

We have, in $E^{0.75 \times 0.25}(s)$ norm,

$$
|\phi_s^i(0^+, s)| + |\phi_u^i(0^-, s)|
$$
$$
\leq C(1 + \frac{1}{|s - \lambda^i(\epsilon)|}) e^{(\sqrt{|s|} + 1)(\eta N - \alpha(L_j^i(\epsilon) - N))}(|\phi_s^i(-L_j^i(\epsilon), s)| + |\phi_u^i(L_j^i(\epsilon), s)|)
$$
$$
\leq C(1 + \frac{1}{|s - \lambda^i(\epsilon)|}) e^{\eta N - \alpha(L_j^i(\epsilon) - N)}|\{\delta^i\}|.
$$

Consider $\phi = T^i(\xi, 0, s)\phi_s^i(0^+, s)$, $\xi \in [0, L_j^i(\epsilon)]$. Using the same exponential dichotomy in $[0, L_j^i(\epsilon)]$, guaranteed by **P4**,

$$
|\phi(\xi, s)| \leq K e^{-\alpha(\sqrt{|s|} + 1)\xi}|\phi_s^i(0^+, s)|, \ \text{Re}(s) > \gamma.
$$

From Lemma 3.1, $u = \mathcal{L}^{-1}\pi_1[P_1\phi]$ is a solution in $H^{2,1}(\gamma)$.

$$
|u|_{H^{2,1}(\gamma)} \leq C|\phi_s^i(0^+, s)|
$$
$$
\leq C(1 + \frac{1}{\sigma - \lambda^i(\epsilon)}) e^{\eta N - \alpha(L_j^i(\epsilon) - N)}|\{\delta^i\}|_{H^{0.75 \times 0.25}(\gamma)}. \tag{4.16}
$$

Similar result holds for $\phi = T^i(\xi, 0, s)\phi_u^i(0^-, x)$, $\xi \in [-L_j^i(\epsilon), 0]$.

For $\text{Re}\, s = \sigma > \gamma$, let

$$
\phi^i = \begin{cases}
T^i(\xi, -L_j^i(\epsilon), s)\phi_s^i(-L_j^i(\epsilon), s) + T^i(\xi, 0, s)\phi_u^i(0^-, s), & \xi \in [-L_j^i(\epsilon), 0], \\
T^i(\xi, 0, s)\phi_s^i(0^+, s) + T^i(\xi, L_j^i(\epsilon), s)\phi_u^i(L_j^i(\epsilon), s), & \xi \in [0, L_j^i(\epsilon)].
\end{cases}
$$

The function $u^i = \mathcal{L}^{-1}\pi_1[P_1\phi^i]$ is a solution for (4.8), but the jump condition (4.10) is not satisfied. In fact, (4.10') has an error

$$
|T^i(L_j^i(\epsilon), 0, s)\phi_s^i(0^+, s) - T^i(-L_j^{i+1}(\epsilon), 0, s)\phi_u^{i+1}(0^-, s)|_{H^{0.75 \times 0.25}(\gamma)}
$$
$$
\leq C(1 + \frac{1}{\gamma - \lambda^i(\epsilon)}) e^{(\eta + \alpha)N - 2\alpha L_j^i(\epsilon)}|\{\delta^i\}|_{H^{0.75 \times 0.25}(\gamma)}.
$$

Due to the fact $\gamma - |\lambda^i(\epsilon)| \geq O_1\epsilon$ and $L_j^i(\epsilon) = \epsilon^{\beta - 1}$ in internal layers, the error is much smaller than $\{\delta^i\}$ if $\epsilon$ is small. It is clear that the desired solution that satisfies both (4.8) and (4.10) can be obtained by iterations. The initial condition (4.9) can be verified by the Paley-Wiener Theorem. The estimates on $|u^i|_{H^{2,1}(\gamma)}$ also follow easily.

We now consider case (2), $0 \leq j \leq r - 1$. The beginning part of the proof is identical to that of case (1). since it is possible that $\gamma < \sup_i(\lambda^i(\epsilon))$, We do not have (4.15) for $\text{Re}(s) > \gamma$. Since $\lambda^i(\epsilon) = O(\epsilon)$, if $\epsilon$ is small, (4.15) is valid for $\text{Re}(s) \geq 1$. It turns out that (4.16) can be replaced by

$$|u|_{H^{2,1}(1)} \leq C(1 + \frac{1}{1 - |\lambda^i(\epsilon)|})e^{\eta N - \alpha(L_j^i(\epsilon) - N)}|\{\delta^i\}|_{H^{0.75 \times 0.25}(\gamma)}.$$

Define $\phi^i$ as in case (1). We need a more precise estimate for $u^i$. In an internal layer, when $\xi \in [-L_j^i(\epsilon), 0]$, we can write $u^i = u_1^i + u_2^i$. Here

$$u_1^i = \mathcal{L}^{-1}[\pi_1(T^i(\xi, -L_j^i(\epsilon), s)\phi_s^i(-L_j^i(\epsilon), s))],$$
$$u_2^i = \mathcal{L}^{-1}[\pi_1(T^i(\xi, 0, s)\phi_u^i(0^-, s))].$$

Similar to case (1), $|u_1^i|_{H^{2,1}(\gamma)} \leq C|\{\delta^i\}|_{H^{0.75 \times 0.25}(\gamma)}$. If $\epsilon$ is small such that $1 - |\lambda^i(\epsilon)| \geq 1/2$, we have

$$|u_2^i|_{H^{2,1}(1)} \leq Ce^{\eta N - \alpha(L_j^i(\epsilon) - N)}|\{\delta^i\}|_{H^{0.75 \times 0.25}(\gamma)}.$$

Using the fact $I_\tau^j = [0, \Delta\tau]$ is a finite interval of length $\Delta\tau$, we have

$$|u_2^i|_{H^{2,1}(\Omega_j^i \times I_\tau^j, \gamma)} \leq e^{(1-\gamma)\Delta\tau}|u_2^i|_{H^{2,1}(\Omega_j^i \times I_\tau^j, 1)}.$$

We have shown

$$|u_2^i|_{H^{2,1}(\Omega_j^i \times I_\tau^j, \gamma)} \leq Ce^{(1-\gamma)\Delta\tau + (\eta+\alpha)N - \alpha L_j^i(\epsilon)}|\{\delta^i\}|_{H^{0.75 \times 0.25}(\gamma)}.$$

By the Trace Theorem, the boundary value of $u_2^i$ at $-L^i(\epsilon)$ satisfies

$$|u_2^i(-L_j^i(\epsilon), \cdot)|_{H^{0.75 \times 0.25}(I_\tau^j, \gamma)} \leq C_1 e^{(1-\gamma)\Delta\tau + (\eta+\alpha)N - \alpha L_j^i(\epsilon)}|\{\delta^i\}|_{H^{0.75 \times 0.25}(\gamma)},$$

Let $0 < \epsilon < \epsilon_0$ be small such that

$$C_1 e^{(1-\gamma)\Delta\tau + (\eta+\alpha)N - \alpha L_j^i(\epsilon)} \leq 1/4.$$

The smallness of $\epsilon_0$ depends on $\Delta\tau$ and $\gamma$. We then have

$$|u_2^i|_{H^{2,1}(\gamma)} \leq C|\{\delta^i\}|_{H^{0.75 \times 0.25}(\gamma)}.$$
$$|u_2^i(-L^i(\epsilon), \cdot)|_{H^{0.75 \times 0.25}(\gamma)} \leq \frac{1}{4}|\{\delta^i\}|_{H^{0.75 \times 0.25}(\gamma)}.$$

Similar result can also be derived for $u^i = u_1^i + u_2^i$ in $[0, L_j^i(\epsilon)]$. Of course, all the above are trivially valid in regular layers.

The rest of the proof is just like that of case (1). $u^i = \mathcal{L}^{-1}[\pi_1\phi^i]$ is a solution to (4.8). The jump condition is not satisfied, but the error is small. The precise solution $\{u^i\}_{-\infty}^\infty$ can be obtained by successive iterations. □

## 5. Corrections in the Final Interval $[t_f, \infty)$

As shown in §4, the original problem is equivalent to the nonlinear system (4.3), (4.6) and (4.7). In this section we study this system in the infinite interval $I_\infty = [t_f, \infty)$. We have shown that (4.3) satisfies **P1** and **P2** in regular layers, and satisfies **P3** and **P4** in internal layers. The nonlinear term $\mathcal{N}^i$ satisfies (4.5). Our main result is the following existence theorem which also contains an estimate on the solution in terms of the initial, residual and jump errors.

Let $\gamma = \sup\{-\epsilon\bar{\gamma}, \dfrac{\lambda^i(\epsilon)}{4}$ for all $i \in \mathbb{Z}\}$ where $\bar{\gamma}$ is the constant in (4.5). Let $g^i \in X(\Omega^i_\infty \times \mathbb{R}^+, \gamma)$, $u^i_0 \in H^1(\Omega^i_\infty)$, $\tau = (t - t_f)/\epsilon$ and $\delta^i \in X^{0.75 \times 0.25}(\gamma)$.

**Theorem 5.1.** *Let $r_1 > 1.5$, $|g^i|_{X(\gamma)} = o(\epsilon^{r_1+1.5})$, $|u^i_0|_{H^1} = o(\epsilon^{r_1+0.5})$, $|\delta^i|_{H^{0.75 \times 0.25}(\gamma)} = o(\epsilon^{r_1})$. Let $t_f$ be $\epsilon$ dependent such that $e^{-\bar{\gamma}t_f} \leq C_0\epsilon^2$, where $\bar{\gamma}$ is the constant in (4.5). Then there exists $\epsilon_0 > 0$ such that for $0 < \epsilon < \epsilon_0$, the nonlinear system (4.3), (4.6) and (4.7) has a unique solution $\{u^i\}$ satisfying the following estimate,*

$$|\{u^i\}|_{X^{2,1}(\gamma)} \leq C(\{\epsilon^{-0.5}|\{u^i_0\}|_{H^1} + |\{\delta^i\}|_{X^{0.75 \times 0.25}(\gamma)} + \epsilon^{-1.5}|\{g^i\}|_{X(\gamma)}\}). \tag{5.1}$$

The proof of Theorem 5.1 is based on the existence of solutions to a linearized problem. Consider,

$$u^i_\tau = u^i_{\xi\xi} + V^i(\xi)u^i_\xi + A^i(\xi)u^i + F^i, \quad \xi \in \Omega^i_\infty,\ \tau > 0, \tag{5.2}$$

$$\begin{pmatrix} I \\ D_\xi \end{pmatrix}(u^i(L^i_\infty(\epsilon), \tau) - u^{i+1}(-L^{i+1}_\infty(\epsilon), \tau)) = \delta^i(\tau), \tag{5.3}$$

$$u^i(\xi, 0) = u^i_0(\xi), \tag{5.4}$$

with the compatibility condition

$$\pi_1\delta^i(0) = u^i_0(L^i_\infty(\epsilon)) - u^{i+1}_0(-L^{i+1}_\infty(\epsilon)).$$

**Lemma 5.2.** *Assume that (5.2) satisfies **P1** and **P2** in regular layers, and satisfies **P3** and **P4** in internal layers. Then there exists $\epsilon_0 > 0$ such that for $0 < \epsilon < \epsilon_0$, system (5.2)–(5.4) has a unique solution $\{u^i\}^\infty_{-\infty}$, $u^i \in X^{2,1}(\Omega^i_\infty \times \mathbb{R}^+, \gamma)$, $i \in \mathbb{Z}$. Also*

$$|u^i|_{X^{2,1}(\gamma)} \leq C\{\epsilon^{-0.5}|\{u^i_0\}|_{H^1} + \epsilon^{-1.5}|\{F^i\}|_{X(\gamma)} + |\{\delta^i\}|_{X^{0.75 \times 0.25}(\gamma)}\}. \tag{5.5}$$

The proof of Lemma 5.2 will be given at the end of this section. We now use Lemma 5.2 to prove Theorem 5.1.

*Proof of Theorem 5.1.* Let the solution in Lemma 5.2 be denoted by

$$\{u^i\} = \mathcal{F}(\{u_0^i\}, \{\delta^i\}, \{F^i\}).$$

We use the Contraction Mapping Theorem to solve (4.3), (4.6) and (4.7). For $\{U^i\} \in X^{2,1}(\gamma)$, let

$$\{\bar{u}^i\} = \mathcal{F}(\{u_0^i\}, \{\delta^i\}, \{\mathcal{N}^i(U^i, \xi, \tau)\}).$$

If $|\{U^i\}|_{X^{2,1}(\gamma)} \leq \epsilon^{r_1}$, then from (4.5), $|\mathcal{N}^i|_{X(\gamma)} = o(\epsilon^{r_1+1.5})$. From Lemma 5.2, if $\epsilon_0$ is small and $0 < \epsilon < \epsilon_0$, $|\{\bar{u}^i\}|_{X^{2,1}(\gamma)} \leq \epsilon^{r_1}$. Thus $\mathcal{F}$ maps a ball of radius $\epsilon^{r_1}$ into itself.

From Lemma 5.2 and the second estimate of (4.5), it is also easy to show that $\mathcal{F}$ is a contraction in such ball. Therefore there exists a unique fixes point $\{u^i\}$ for $\mathcal{F}$. Using Lemma 5.2 again, we have (5.1). □

*Proof of Lemma 5.2.* There exist bounded extensions of $u_0^i$, $F^i$ to $\xi \in \mathbb{R}$. Without loss of generality, assume that $u_0^i \in H^1(\mathbb{R})$, $F^i \in X(\mathbb{R} \times \mathbb{R}^+, \gamma)$.

(I) First, consider (5.2) and (5.4) for $\xi \in \mathbb{R}$, and ignore the boundary conditions (5.3). The solution can be written as

$$u^i(\tau) = e^{\mathcal{A}^i\tau}u_0^i + \int_0^\tau e^{\mathcal{A}^i(\tau-s)}F^i(s)ds.$$

In regular layers, using Lemma 3.11 with $\alpha_0 = -\bar{\sigma}/2$, it is easy to see

$$|u^i|_{X^{2,1}(\gamma)} \leq C(|u_0^i|_{H^1} + |F^i|_{X(\gamma)}).$$

In an internal layer, we make the following spectral decompositions.

$$u_0^i = \alpha^i\phi^i + u_0^{i\perp},$$
$$F^i(s) = \beta^i(s)\phi^i + F^{i\perp}(s),$$
$$u^i(\tau) = u_1^i(\tau) + u_2^i(\tau).$$

Here $\phi^i$ is the eigenvector associated to the eigenvalue $\lambda^i(\epsilon)$, $u_1^i(\tau)$ is in the eigenspace spanned by $\phi^i$, the functions $u_2^i(\tau)$, $u_0^{i\perp}$ and $F^{i\perp}(s)$ are in the spectral space corresponding to $\text{Re}\lambda \leq -\bar{\sigma} < 0$.

$$u_2^i(\tau) = e^{\mathcal{A}^i\tau}u_0^{i\perp} + \int_0^\tau e^{\mathcal{A}^i(\tau-s)}F^{i\perp}(s)ds.$$

Using Lemma 3.11 with $\alpha_0 = -\bar{\sigma}/2$, it is easy to see

$$|u_2^i|_{X^{2,1}(\gamma)} \leq C(|u_0^{i\perp}|_{H^1} + |F^{i\perp}|_{X(\gamma)}).$$

$$u_1^i(\tau) = \alpha^i e^{\lambda^i(\epsilon)\tau}\phi^i + (\int_0^\tau e^{\lambda^i(\epsilon)(\tau-s)}\beta^i(s)ds)\phi^i = I_1 + I_2.$$

Using the fact $\dfrac{\lambda^i(\epsilon)}{4} \leq \gamma < 0$, we can verify that

$$|e^{-\gamma\tau}I_1|_{H^{2,1}} \leq C|\alpha^i||\phi^i|_{H^2}/\sqrt{|\gamma|}.$$

Let $\Delta\beta^i(s) = \beta^i(s) - \beta^i(\infty)$. Then $e^{-\gamma\tau}\Delta\beta^i \in L^2(\mathbb{R}^+)$.

$$\begin{aligned}
I_2 &= [\int_0^\tau e^{\lambda^i(\epsilon)(\tau-s)}(\beta^i(\infty) + \Delta\beta^i(s))ds]\phi^i \\
&= \frac{\beta^i(\infty)}{\lambda^i(\epsilon)}\phi^i - \frac{\beta^i(\infty)}{\lambda^i(\epsilon)}e^{\lambda^i(\epsilon)\tau}\phi^i + [e^{\lambda^i(\epsilon)\tau} * \Delta\beta^i(\tau)]\phi^i \\
&= J_1 + J_2 + J_3.
\end{aligned}$$

The first term is time independent with $|J_1|_{H^2} \leq \dfrac{C}{\epsilon}|F^i|_{X(\gamma)}$. Since $|e^{-\gamma\tau}e^{\lambda^i(\epsilon)\tau}|_{L^2} \leq 1/\sqrt{|\gamma|}$,

$$\begin{aligned}
|J_2|_{X^{2,1}(\gamma)} &\leq \frac{C}{|\gamma|^{1.5}}|\beta^i(\infty)||\phi^i|_{H^2} \\
&\leq \frac{C}{|\gamma|^{1.5}}|F^i|_{X(\gamma)}.
\end{aligned}$$

Since

$$\begin{aligned}
|e^{-\gamma\tau}J_3|_{L^2} &= |(e^{(\lambda^i(\epsilon)-\gamma)\tau}) * (e^{-\gamma\tau}\Delta\beta^i(\tau))|_{L^2} \\
&\leq |(e^{(\lambda^i(\epsilon)-\gamma)\tau})|_{L^1}|(e^{-\gamma\tau}\Delta\beta^i(\tau))|_{L^2} \\
&\leq \frac{C}{|\gamma|}|\Delta\beta^i|_{L^2(\gamma)},
\end{aligned}$$

we have

$$|J_3|_{H^{2,1}(\gamma)} \leq \frac{C}{|\gamma|}|F^i|_{X(\gamma)}.$$

Therefore

$$|u_1^i|_{X^{2,1}(\gamma)} \leq C(\epsilon^{-0.5}|u_0^i|_{H^1} + \epsilon^{-1.5}|F^i|_{X(\gamma)}).$$

Let the solution of (5.2) and (5.4) be denoted by $\bar{u}^i = u_1^i + u_2^i$. We have shown

$$|\bar{u}^i|_{X^{2,1}(\gamma)} \leq C(\epsilon^{-0.5}|u_0^i|_{H^1} + \epsilon^{-1.5}|F^i|_{X(\gamma)}).$$

We now restrict $\bar{u}^i$ to the domain $\Omega_\infty^i = [-L_\infty^i(\epsilon), L_\infty^i(\epsilon)]$. At the common boundary of $\Omega_\infty^i$ and $\Omega_\infty^{i+1}$ there is a jump $\bar{\delta}^i = \begin{pmatrix} I \\ D_\xi \end{pmatrix}(\bar{u}^i(L_\infty^i(\epsilon)) - \bar{u}^{i+1}(-L_\infty^{i+1}(\epsilon))$. We can show that

$$|\bar{\delta}^i|_{X^{0.75 \times 0.25}} \leq C(|u_0^i|_{H^1} + |F^i|_{X(\gamma)}). \tag{5.6}$$

In fact, in regular layers, using the Trace Theorem,

$$\left|\begin{pmatrix} I \\ D_\xi \end{pmatrix}(u^i(\pm L_\infty^i(\epsilon)))\right| \leq C(|u_0^i|_{H^1} + |F^i|_{X(\gamma)}).$$

Based on the Trace Theorem again, the above estimate is also valid for $u_2^i$ in internal layers. For $u_1^i$ in the internal layer, let

$$u_1^i(\tau) = h^i(\tau)\phi^i.$$

$$\begin{pmatrix} I \\ D_\xi \end{pmatrix}u_1^i(\tau) = h^i(\tau)\begin{pmatrix} \phi^i \\ \phi_\xi^i \end{pmatrix}.$$

Here $|h^i(\tau)|_{H^1(\mathbb{R}^+)} \leq C\epsilon^{-1.5}(|u_0^i|_{H^1} + |F^i|_{X(\gamma)})$. At the boundaries, $|\xi| = \epsilon^{\beta-1}$, $|\phi^i|_{H^{0.75}}$
$+ |\phi_\xi^i|_{H^{0.25}} \leq Ce^{-\alpha\epsilon^{\beta-1}} \leq C\epsilon^{1.5}$, if $\epsilon$ is sufficiently small. Thus

$$\left|\begin{pmatrix} I \\ D_\xi \end{pmatrix}(u_1^i(\pm\epsilon^{\beta-1}))\right|_{X^{0.75\times0.25}(\gamma)} \leq |h^i|_{H^1}(|\phi^i(\pm\epsilon^{\beta-1})| + |\phi_\xi^i(\pm\epsilon^{\beta-1})|)$$
$$\leq C(|u_0^i|_{H^1} + |F^i|_{X(\gamma)}).$$

This proves (5.6), since $\bar{u}^i = u_1^i + u_2^i$.

(II) Consider a boundary value problem with zero initial condition and zero nonhomogeneous term,

$$u_\tau^i = u_{\xi\xi}^i + V^i(\xi)u_\xi^i + A^i(\xi)u^i, \tag{5.7}$$

$$u^i(0) = 0, \tag{5.8}$$

$$\tilde{\delta}^i = \begin{pmatrix} I \\ D_\xi \end{pmatrix}(u^i(L_\infty^i(\epsilon)) - u^{i+1}(-L_\infty^{i+1}(\epsilon))) = \delta^i - \bar{\delta}^i. \tag{5.9}$$

It is obvious that $\pi_1\tilde{\delta}^i(0) = \pi_1\delta^i(0) - \pi_1\bar{\delta}^i(0) = 0$. Also, due to (5.6),

$$|\tilde{\delta}^i|_{X^{0.75\times0.25}(\gamma)} \leq |\delta^i|_{X^{0.75\times0.25}(\gamma)} + C(|\{u_0^i\}|_{H^1} + |\{F^i\}|_{X(\gamma)}). \tag{5.10}$$

Let the solution to (5.7)–(5.9) be $\{\bar{u}^i\}_{-\infty}^\infty$. We want to show that

$$|\bar{u}^i|_{X^{2,1}(\gamma)} \leq C|\{\tilde{\delta}^i\}|_{X^{0.75\times0.25}(\gamma)}. \tag{5.11}$$

Then from (5.10), and the fact $u^i = \bar{u}^i + \bar{\bar{u}}^i$ is a solution to (5.2)–(5.4), we will have proved the theorem.

To prove (5.11), we write the solution as $\bar{\bar{u}}^i = \tilde{u}_1^i + \tilde{u}_2^i$. The function $\tilde{u}_1^i$ satisfies a time independent system,

$$u_{\xi\xi}^i + V^i(\xi)u_\xi^i + A^i(\xi)u^i = 0,$$

$$\begin{pmatrix} I \\ D_\xi \end{pmatrix}(u^i(L_\infty^i(\epsilon)) - u^{i+1}(-L_\infty^{i+1}(\epsilon))) = \tilde{\delta}^i(\infty) \in \mathbb{R}^{2n}.$$

The solution for that system uniquely exists and satisfies

$$|\tilde{u}_1^i|_{H^2} \le C|\{\tilde{\delta}^i(\infty)\}|_{\mathbb{R}^{2n}}. \tag{5.12}$$

By converting the elliptic equation into a first order system of ODEs, the proof can be derived from [23], thus, will be skipped here.

The function $\tilde{u}_2^i$ satisfies (5.7) with

$$\binom{I}{D_\xi}(\tilde{u}_2^i(L_\infty^i(\epsilon)) - \tilde{u}_2^{i+1}(-L_\infty^{i+1}(\epsilon))) = d^i, \tag{5.13}$$

$$\tilde{u}_2^i(\xi,0) = -\tilde{u}_1^i(\xi). \tag{5.14}$$

Here $d^i(\tau) = \tilde{\delta}^i(\tau) - \tilde{\delta}^i(\infty) \in H^{0.75 \times 0.25}(\gamma)$. By (5.14), the compatibility condition

$$\pi_1 d^i(0) = -\pi_1 \tilde{\delta}^i(\infty)$$
$$= \tilde{u}_2^i(L^i(\epsilon),0) - \tilde{u}_2^{i+1}(-L^{i+1}(\epsilon),0),$$

is clearly satisfied.

To find $\tilde{u}_2^i$, let $\tilde{u}_2^i = u_3^i + u_4^i$. We want $u_3^i$ to satisfy the initial condition (5.14). Therefore, $u_4^i$ has zero initial condition, so that Lemma 4.1 can be applied on $\{u_4^i\}$. Assume that $\tilde{u}_1^i$ has a bounded extension to $H^1(\mathbb{R})$. Let $Q_0$ and $Q_s$ be the spectral projections associated to $\lambda^i(\epsilon)$ and $\text{Re}(\lambda) \le -\bar{\sigma}$. Let

$$u_3^i = Q_0 u_3^i + Q_s u_3^i$$
$$= e^{\mathcal{A}^i \tau}(-Q_0 \tilde{u}_1^i) + e^{\mathcal{A}^i \tau}(-Q_s \tilde{u}_1^i).$$

Let $Q_0 u_3^i(\tau) = h_3^i(\tau)\phi^i$ where $\phi^i$ is the eigenvector, corresponding to $\lambda^i(\epsilon)$.

$$|h_3^i(\tau)|_{L^2} \le C|e^{\lambda^i(\epsilon)\tau} h_3^i(0)|_{L^2} \le \frac{C}{\sqrt{\epsilon}}|h_3^i(0)|,$$

where $|h_3^i(0)| \le C|Q_0 \tilde{u}_1^i|_{H^1}$. Recall that the spectrum of $\mathcal{A}^i$ in $\mathcal{R}Q_s$ is in $\text{Re}\lambda \le -\bar{\sigma}$. We have

$$|u_3^i|_{H^{2,1}(\gamma)} \le \frac{C_1}{\sqrt{\epsilon}}|Q_0 \tilde{u}_1^i|_{H^1} + C_2|Q_s \tilde{u}_1^i|_{H^1}. \tag{5.15}$$

Restrict $u_3^i$ to the domain $\Omega_\infty^i = [-L_\infty^i(\epsilon), L_\infty^i(\epsilon)]$. At the boundaries, there is a small jump,

$$\binom{I}{D_\xi}(u_3^i(L_\infty^i(\epsilon),\tau) - u^{i+1}(-L_\infty^{i+1}(\epsilon),\tau)) = d_3^i(\tau).$$

We can show that

$$|d_3^i|_{H^{0.75 \times 0.25}} \le C|\{\tilde{u}_1^i\}|_{H^1} \le C|\{\tilde{\delta}^i(\infty)\}|_{\mathbb{R}^{2n}}. \tag{5.16}$$

In fact, the traces of $Q_s u_3^i$ at $\pm L_\infty^i(\epsilon)$ is bounded by $|Q_s \tilde{u}_1^i|_{H^1}$. And using the exponential decay of $\phi^i$ and $\phi_\xi^i$ as $|\xi| \to \infty$, the trace of $Q_0 u_3^i$ at $\pm L_\infty^i(\epsilon)$ is

$$|h_3^i(\tau)\phi^i(\pm\epsilon^{\beta-1})|_{H^{0.75 \times 0.25}(\gamma)} \leq C|h_3^i|_{H^1}(|\phi^i(\pm\epsilon^{\beta-1})| + |\phi_\xi^i(\pm\epsilon^{\beta-1})|)$$
$$\leq C|Q_0 \tilde{u}_1^i|_{H^1}.$$

Thus at $\pm L_\infty^i(\epsilon)$, the traces of $Q_0 u_3^i(\xi)$ are also bounded by $C|\{\tilde{u}_1^i\}|_{H^1}$. This proves (5.16).

We now look for $u_4^i$ that satisfies (5.7), (5.8), with

$$\binom{I}{D_\xi}(u_4^i(L^i(\epsilon), \tau) - u^{i+1}(L^{i+1}(\epsilon), \tau)) = d^i(\tau) - d_3^i(\tau) = D^i(\tau). \tag{5.17}$$

Observe that

$$\pi_1 d^i(0) = -\pi_1 \tilde{\delta}^i(\infty),$$
$$\pi_1 d_3^i(0) = u_3^i(L_\infty^i(\epsilon), 0) - u_3^{i+1}(-L^{i+1}(\epsilon), 0)$$
$$= -(\tilde{u}_1^i(L_\infty^i(\epsilon)) - \tilde{u}_1^{i+1}(-L_\infty^{i+1}(\epsilon)))$$
$$= -\pi_1 \tilde{\delta}^i(\infty).$$

Therefore, the compatibility condition $\pi_1 D^i(0) = 0$ is valid. Moreover,

$$|D^i(\tau)|_{H^{0.75 \times 0.25}(\gamma)} \leq |d^i|_{H^{0.75 \times 0.25}(\gamma)} + |d_3^i|_{H^{0.75 \times 0.25}(\gamma)}$$
$$\leq C|\tilde{\delta}^i|_{X^{0.75 \times 0.25}(\gamma)}.$$

Using Lemma 4.1 to (5.7), (5.8) and (5.17), we have a unique solution $\{u_4^i\}$ with

$$|u_4^i|_{H^{2,1}(\gamma)} \leq C|\{D^i\}|_{H^{0.75 \times 0.25}(\gamma)} \tag{5.18}$$
$$\leq |\{\tilde{\delta}^i\}|_{H^{0.75 \times 0.25}(\gamma)}.$$

Recall that $\bar{\bar{u}}^i = \tilde{u}_1^i + \tilde{u}_2^i$, and $\tilde{u}_2^i = u_3^i + u_4^i$. The part (II) of the proof is completed by combining (5.12), (5.15) and (5.18). □

## 6. Correction in Finite Intervals

The main results of this section are Theorems 6.1 and 6.2.

Recall from §5, in order to use Theorem 5.1 in $[t_f, \infty)$, $t_f \geq \log(1/(C_0\epsilon^2))/\bar{\gamma}$. Define

$$r = [\log(1/(C_0\epsilon^2))/\epsilon\bar{\gamma}\Delta\tau] + 1,$$
$$\Delta t = \epsilon\Delta\tau, \quad t_f = r\Delta t,$$
$$I_j = [j\Delta t, (j+1)\Delta t], \, 0 \leq j \leq r-1, \quad I_\infty = [t_f, \infty), \tag{6.1}$$

where $[x]$ is the integer part of $x$. As in §4, the coordinate change $\mathfrak{R}(0) \to \mathfrak{R}^i(1)$ maps $\Sigma^i \times I_j$ to $\Omega_j^i \times I_\tau^j$. The solutions of (1.1) in $I_j$ are equivalent to the solutions of (4.3), (4.6) and (4.7) in intervals $I_\tau^j$. More precisely, the solutions are indexed by $j$ and will be denoted by $u_j^i$. The initial data, jumps and forcing terms are $u_j^i(0)$, $\delta_j^i$, $g_j^i$ and the coefficients are $A_j^i(\xi)$, $V_j^i(\xi)$. The domain for equation (4.3) is $\Omega_j^i \times I_\tau^j$. Recall that $I_\tau^j = [0, \Delta\tau]$, $0 \le j \le r - 1$, and $I_\tau^\infty = \mathbb{R}^+$, and $\Omega_j^i = (-L_j^i(\epsilon), L_j^i(\epsilon))$ where $L_j^i(\epsilon) = \Delta y^i(\bar{t})/2\epsilon$, $\bar{t} = j\Delta t$.

In the new notations, consider,

$$u_{j\tau}^i = u_{j\xi\xi}^i + V_j^i(\xi)u_{j\xi}^i + A_j^i(\xi)u_j^i + \mathcal{N}_j^i(u_j^i, \xi, t, \epsilon), \quad \xi \in \Omega_j^i, \ \tau \in I_\tau^j, \tag{6.2}$$

$$\binom{I}{D_\xi} [u_j^i(L_j^i(\epsilon), \tau, \epsilon) - u_j^{i+1}(-L_j^{i+1}(\epsilon), \tau, \epsilon)] = \delta_j^i(\tau), \tag{6.3}$$

$$u_{j+1}^i(\xi, 0) = u_j^i(\xi_1, \Delta\tau), \ u_0^i(\xi, 0) = u_0^i(\xi). \tag{6.4}$$

In (6.4), it is understood that $u_r^i = u_\infty^i$, and $(\xi, \tau) \to (\xi_1, \tau_1)$ is the change of coordinates $\mathfrak{R}^i(1) \to \mathfrak{R}(0) \to \mathfrak{R}^i(1)$ that maps

$$\Omega_{j+1}^i \times I_\tau^{j+1} \to \Sigma^i \cap I_{j+1} \to \Sigma^i \cap I_j \to \Omega_j^i \times I_\tau^j.$$

With $\bar{t} = j\Delta t$, define,

$$\xi \xrightarrow{\nu_1} x \xrightarrow{\nu_2} y \xrightarrow{\nu_3} \xi_1,$$

$$\nu_1: \quad x = m^i(\bar{t} + \Delta t) + \epsilon\xi,$$

$$\nu_2: \quad y = x - \Xi(x, \bar{t} + \Delta t, \bar{t}, i),$$

$$\nu_3: \quad \xi_1 = (y - m^i(\bar{t}))/\epsilon.$$

We have

$$u_{j+1}^i(\xi, 0) = u_j^i(\nu_3 \circ \nu_2 \circ \nu_1(\xi), \Delta\tau).$$

We have showed, in §4, that (4.3) satisfies **P1** and **P2** in regular layers, and satisfies **P3** and **P4** in internal layers. Also $\mathcal{N}_j^i$ satisfies (4.4). In each finite time interval, the following existence of solutions holds.

**Theorem 6.1.** *For any $\Delta\tau > 0$, let the intervals $I_j$, $0 \le j \le r - 1$, be constructed by (6.1). Consider (6.2) and (6.3) in an time interval $I_\tau^j$—the image of $I_j$ by the change of variable $\tau = (t - j\Delta t)/\epsilon$. Assume that $g_j^i \in L^2(\Omega_j^i \times I_\tau^j)$, $u_j^i(0) \in H^1(\Omega_j^i)$, and $\delta_j^i \in H^{0.75 \times 0.25}(I_\tau^j)$, with $|\{g_j^i\}|_{L^2} = o(\epsilon)$, $|\{u_j^i(0)\}|_{H^1} = o(\epsilon)$, $|\{\delta_j^i\}|_{H^{0.75.25}} = o(\epsilon)$. Then there exists $\epsilon_0 > 0$ such that for $0 < \epsilon < \epsilon_0$, (6.2)–(6.3) has a unique solution $\{u_j^i\}_{-\infty}^\infty$, $u_j^i \in H^{2,1}(\Omega_j^i \times I_\tau^j)$. Also*

$$|u_j^i|_{H^{2,1}} \le Ce^{\mu\Delta\tau}\Delta\tau(|\{u_j^i(0)\}|_{H^1} + |\{g_j^i\}|_{L^2}) + C|\{\delta_j^i\}|_{H^{0.75 \times 0.25}(I_\tau^j)}.$$

In the time scale $t$, the length of the interval $[0, \Delta\tau]$ is only $\epsilon\Delta\tau = O(\epsilon)$. Since $u^i_j(\Delta\tau)$, after a near identity change of variable, is the initial data for the next time interval, and since $r \to \infty$ as $\epsilon \to 0$, the result in Theorem 6.1 is not accurate enough to guarantee the existence of solutions in $I_\infty$. We will prove a sharper estimate in Lemma 6.4 showing that

$$|u^i_j(\Delta\tau)|_{H^1} \leq (1 + C\epsilon)|\{u^i_j(0)\}|_{H^1} + \dots . \tag{6.5}$$

With this, even if $r = O(-\log(\epsilon)/\epsilon)$, using $(1 + C\epsilon)^{1/C\epsilon} \leq e$. the accumulation of error will not grow too fast as $\epsilon \to 0$. We show in Theorem 6.2, that the accumulation error at $t_f = r\Delta t$ is small enough so that the existence of solutions in $I_j$ is guaranteed by Theorem 6.1, and in $I_\infty$ is guaranteed by Theorem 5.1. To this end, let us reconsider (6.2)–(6.4) with an emphasis on the norm $|u^i_j(\Delta\tau)|_{H^1}$ in term of $|\{u^i_j(0)\}|_{H^1}$.

**Theorem 6.2.** *There exist $\Delta\tau > 0$ such that if $I_j$ and $I_\infty$ are defined by (6.1) using this $\Delta\tau$, and then we have the following properties concerning the solutions of (6.2)–(6.4). If $r_1 > 1.5$, $J_1 = \frac{2C}{\bar{\gamma}\Delta\tau}$, where $C$ is from (6.5) and $\bar{\gamma}$ from Theorem 5.1, $|\{u^i_0(0)\}|_{H^1} = o(\epsilon^{r_1 + 0.5 + J_1})$, $|\{\delta^i_j\}|_{H^{0.75 \times 0.25}} + |\{g^i_j\}|_{L^2} = o(\epsilon^{r_1 + 1.5 + J_1})$, then system (6.2)–(6.4), $0 \leq j \leq r - 1$, has a unique solution for $0 < \epsilon < \epsilon_0$, where $\epsilon_0 > 0$ is a small constant. The solution in each time interval satisfies*

$$|\{u^i_j(0)\}|_{H^1(\Omega^i_j)} \leq C\epsilon^{-J_1}|\{u^i_0\}|_{H^1(\Omega^i_0)} + C\epsilon^{-J_1 - 1}\sup_{\nu < j}(|\{\delta^i_\nu\}|_{H^{0.75 \times 0.25}} + |\{g^i_\nu\}|_{L^2}).$$

*In particular, $|\{u^i_\infty(0)\}|_{H^1} = o(\epsilon^{r_1 + 0.5})$.*

The proofs of Theorems 6.1 and 6.2 are based on Lemmas 6.3 and 6.4. We present these lemmas first.

Since all $I^j_\tau$, $0 \leq j \leq r - 1$, are identical, they are denoted by $I_\tau$ for simplicity. Assume $\Delta\tau \geq 1$, in $I_\tau = [0, \Delta\tau]$. Consider the following initial value problem,

$$u_\tau = u_{\xi\xi} + V(\xi)u_\xi + A(\xi)u + F(\xi, \tau), \quad \xi \in \mathbb{R}, \, \tau \in I_\tau, \tag{6.6}$$

$$u(\xi, 0) = u_0(\xi). \tag{6.7}$$

**Lemma 6.3.** *Assume that the linear operator $\mathcal{A} : L^2 \to L^2$, $u \to u_{\xi\xi} + V(\xi)u_\xi + A(\xi)u$ has a simple eigenvalue $\lambda_0$ that is close to zero, all the other spectra are contained in $\{Re\lambda \leq -\bar{\sigma}\}$ for some constant $\bar{\sigma} > 0$. Let $-\bar{\sigma} < \gamma \leq 0$, and $\mu = \max(0, Re\lambda_0)$. Let $F \in L^2(\mathbb{R} \times I_\tau)$ and $u_0 \in H^1(\mathbb{R})$. Then the solution of (6.6), (6.7) satisfies*

$$|u|_{H^{2,1}(\mathbb{R} \times I_\tau)} \leq Ce^{\mu\Delta\tau}\Delta\tau(|u_0|_{H^1} + |F|_{L^2}). \tag{6.8}$$

*Moreover, let $Q_0$ and $Q_s$ be the spectral projection corresponding to $\lambda_0$ and the spectra in $\{Re\lambda \leq -\bar{\sigma}\}$. Since $u : I_\tau \to L^2(\mathbb{R})$ is continuous, we can write $u(\tau) = h(\tau)\phi + Q_s u(\tau)$, where $h(\tau)\phi = Q_0 u(\tau)$. Then*

*(a)* $|h(\tau)| \leq e^{Re\lambda_0\tau}|h(0)| + e^{\mu\tau}\sqrt{\tau}|F|_{L^2}$.

*(b)* $|Q_s u(\tau)|_{H^1} \leq C(e^{\gamma\tau}|Q_s u_0|_{H^1} + |F|_{L^2})$.

*Proof.* Let $u_0 = h(0)\phi + Q_s u_0$ and $F(\zeta) = \beta(\zeta)\phi + F_s(\zeta)$ be the spectral decompositions corresponding to $Q_0$ and $Q_s$. Applying $Q_s$ to the variation of constant formula, we have

$$Q_s u(\tau) = e^{\mathcal{A}\tau} Q_s u_0 + \int_0^\tau e^{\mathcal{A}(\tau-\zeta)} F_s(\zeta) d\zeta,$$

When restricted to $\mathcal{R}Q_s$, $\mathcal{A}$ is a sectorial operator with $Re\sigma(\mathcal{A}) \leq -\bar{\sigma} < 0$. Thus, $e^{\mathcal{A}\tau}$ is exponentially stable on $\mathcal{R}Q_s$. Therefore, we have (b). And also

$$|Q_s u|_{H^{2,1}(\mathbb{R}\times I_\tau)} \leq C(|u_0|_{H^1} + |F|_{L^2}). \tag{6.9}$$

Using $Q_0$ to the variation of constant formula, we have

$$h(\tau) = e^{\lambda_0\tau} h(0) + \int_0^\tau e^{\lambda_0(\tau-\zeta)}\beta(\zeta)d\zeta.$$

$$|\int_0^\tau e^{\lambda_0(\tau-\zeta)}\beta(\zeta)d\zeta| \leq [\int_0^\tau e^{2\mu\zeta} d\zeta]^{1/2}[\int_0^\tau \beta^2(\zeta)d\zeta]^{1/2}$$
$$\leq e^{\mu\tau}\sqrt{\tau}|\beta|_{L^2(I_\tau)}$$
$$\leq Ce^{\mu\tau}\sqrt{\tau}|F|_{L^2(\mathbb{R}\times I_\tau)}.$$

Observe also that $|e^{\lambda_0\tau} h(0)| \leq e^{Re\lambda_0\tau}|h(0)|$. We have proved (a).

Based on (a), it is elementary to show

$$|h|_{H^1(I_\tau)} \leq Ce^{\mu\Delta\tau}(\sqrt{\Delta\tau}|h(0)| + \Delta\tau|F|_{L^2}). \tag{6.10}$$

Since $|h(0)| \leq C|u_0|_{H^1}$ and $|\beta|_{L^2(I_\tau)} \leq C|F|_{L^2(\mathbb{R}\times I_\tau)}$,

$$|h\phi|_{H^{2,1}(\mathbb{R}\times I_\tau)} \leq C|h|_{H^1(I_\tau)}|\phi|_{H^2}$$
$$\leq Ce^{\mu\Delta\tau}(\sqrt{\Delta\tau}|u_0|_{H^1} + \Delta\tau|F|_{L^2}).$$

Recall that $\Delta\tau \geq 1$. Thus

$$|h\phi|_{H^{2,1}(\mathbb{R}\times I_\tau)} \leq Ce^{\mu\Delta\tau}\Delta\tau(|u_0|_{H^1} + |F|_{L^2}).$$

(6.8) follows from the above and (6.9). $\qquad\qquad\qquad\square$

Consider a linear initial boundary value problem in one of the finite time intervals. Let $I_\tau = [0, \Delta\tau]$ and $\Omega_j^i = (-L_j^i(\epsilon), L_j^i(\epsilon))$.

$$u_\tau^i = u_{\xi\xi}^i + V^i(\xi)u_\xi + A^i(\xi)u + F^i(\xi,\tau), \quad \xi \in \Omega_j^i, \tau \in I_\tau, \tag{6.11}$$

$$\begin{pmatrix} I \\ D_\xi \end{pmatrix} (u^i(L^i(\epsilon)) - u^{i+1}(-L^{i+1}(\epsilon))) = \delta^i, \tag{6.12}$$

$$u^i(\xi, 0) = u_0^i(\xi), \tag{6.13}$$

with the compatibility condition,

$$\pi_1 \delta^i(0) = u_0^i(L^i(\epsilon)) - u_0^{i+1}(-L^{i+1}(\epsilon)).$$

Assume that the coefficients of (6.11) have been extended to $\xi \in \mathbb{R}$ by constants. Let the operator on $u^i$ defined by the right hand side of (6.11) be denoted by $\mathcal{A}^i : L^2(\mathbb{R}) \to L^2(\mathbb{R})$. Let $u_0^i \in H^1(\mathbb{R})$, $F^i \in L^2(\mathbb{R} \times I_\tau)$ and $\delta^i \in H^{0.75 \times 0.25}(I_\tau)$.

Assume that (6.11) satisfies **P1** and **P2** in regular layers and satisfies **P3** and **P4** in internal layers. In internal layers, for any $u \in H^1(\mathbb{R})$, let $u = Q_0 u + Q_s u$ be the spectral decomposition. Let $Q_0 u = h\phi^i$ where $\phi^i$ is the eigenvector corresponding to $\lambda^i(\epsilon)$. Then $|h| + |Q_s u|_{H^1}$ is an equivalent norm to $|u|_{H^1}$. In regular layers, for convenience, let $Q_s u = u$, $h = 0$. Formally, we still have the above equivalent norm. Let $\mu = \max\{0, \sup_i [\text{Re}\lambda^i(\epsilon)]\}$.

**Lemma 6.4.** *There exists $\epsilon_0 > 0$ such that for $0 < \epsilon < \epsilon_0$, (6.11)–(6.13) has a unique solution $\{u^i\}_{-\infty}^\infty$, $u^i \in H^{2,1}(\Omega_j^i \times I_\tau)$ for all $i \in \mathbb{Z}$. Also*

$$|u^i|_{H^{2,1}} \leq Ce^{\mu\Delta\tau}\Delta\tau(|\{u_0^i\}|_{H^1} + |\{F^i\}|_{L^2}) + C|\{\delta^i\}|_{H^{0.75 \times 0.25}(I_\tau)}. \tag{6.14}$$

*Moreover, if $\Delta\tau$ is sufficiently large, then the solution admits a bounded extension to $\xi \in \mathbb{R}$ such that if $Q_0 u^i = h^i \phi^i$ then,*

$$|h^i(\Delta\tau)| + |Q_s u^i(\Delta\tau)|_{H^1(\mathbb{R})} \leq (1 + C\epsilon) \sup_i (|h^i(0)| + |Q_s u^i(0)|_{H^1}) \tag{6.15}$$

$$+ C|\{\delta^i\}|_{H^{0.75 \times 0.25}(I_\tau)} + C|\{F^i\}|_{L^2}.$$

*The constant $C$ depends on $\Delta\tau$, but not $\epsilon$.*

*Proof.* The solution will be written as $u^i = \bar{u}^i + \bar{\bar{u}}^i$. Using the spectral projections,

$$u^i(\tau) = h^i(\tau)\phi^i + Q_s u^i(\tau),$$

$$\bar{u}^i(\tau) = \bar{h}^i(\tau)\phi^i + Q_s \bar{u}^i(\tau),$$

$$\bar{\bar{u}}^i(\tau) = \bar{\bar{h}}^i(\tau)\phi^i + Q_s \bar{\bar{u}}^i(\tau).$$

Let $\bar{u}^i(\tau) = e^{\mathcal{A}^i \tau} u_0^i + \int_0^\tau e^{\mathcal{A}^i(\tau-s)} F^i(s) ds$. In internal layers, using Lemma 6.3, (6.8), we have

$$|\bar{u}^i|_{H^{2,1}(\mathbb{R}\times I_\tau)} \leq C e^{\mu\Delta\tau} \Delta\tau (|u_0^i|_{H^1} + |F^i|_{L^2}). \tag{6.16}$$

(6.16) is also valid in regular layers.

In internal layers, $\bar{u}^i(\tau) = \bar{h}^i(\tau)\phi^i + Q_s\bar{u}^i(\tau)$. Note that $\bar{u}^i(0) = u_0^i$, $\bar{h}^i(0) = h^i(0)$, $Q_s\bar{u}^i(0) = Q_s u_0^i$. We have, using Lemma 6.3 (a) and (b), for some $-\bar{\sigma}/4 < \gamma \leq 0$,

$$\begin{aligned} |\bar{h}^i(\Delta\tau)| &\leq e^{\mu\Delta\tau}(|h^i(0)| + \sqrt{\Delta\tau}|F^i|_{L^2}), \\ |Q_s\bar{u}^i(\Delta\tau)|_{H^1} &\leq C[e^{\gamma\Delta\tau}|Q_s u_0^i|_{H^1} + |F^i|_{L^2}]. \end{aligned} \tag{6.17}$$

(6.17) is also valid in regular layers, with $h^i = 0$.

At $\xi = \pm L_j^i(\epsilon)$,

$$|\bar{u}^i(\pm L_j^i(\epsilon), \tau)| \leq |\bar{h}^i(\tau)||\phi^i(\pm L_j^i(\epsilon))| + |Q_s\bar{u}^i(\pm L_j^i(\epsilon), \tau)|.$$

It is clear that from the Trace Theorem,

$$\begin{aligned} |\begin{pmatrix} I \\ D_\xi \end{pmatrix} Q_s\bar{u}^i(\pm L_j^i(\epsilon), \cdot)|_{H^{0.75\times0.25}(\gamma)} &\leq C|Q_s\bar{u}^i|_{H^{2,1}(\gamma)} \\ &\leq C(|Q_s u_0^i|_{H^1(\mathbb{R})} + e^{|\gamma|\Delta\tau}|F^i|_{L^2}). \end{aligned}$$

The second inequality is based on Lemma 3.11.

Though $Q_0\bar{u}^i$ does not decay exponentially in time, we can still consider weighted norm in the finite interval $I_\tau$. Observe that

$$|\bar{h}^i|_{H^1(I_\tau, \gamma)} \leq C e^{|\gamma|\Delta\tau} |\bar{h}^i|_{H^1(I_\tau)}.$$

Using (6.10) and $|\phi^i(\xi)| + |\partial_\xi \phi^i(\xi)| \leq C e^{\gamma_1|\xi|}$ for some $\gamma_1 < 0$, we have

$$|\begin{pmatrix} I \\ D_\xi \end{pmatrix} Q_0\bar{u}^i(\pm L_j^i(\epsilon), \cdot)|_{H^{0.75\times0.25}(\gamma)} \leq C e^{(\mu+|\gamma|)\Delta\tau} e^{\gamma_1 \epsilon^{\beta-1}} (\sqrt{\Delta\tau}|h^i(0)| + \Delta\tau |F^i|_{L^2}).$$

Let $\epsilon_0 > 0$ be small so that $C e^{(\mu+|\gamma|)\Delta\tau + \gamma_1 \epsilon^{\beta-1}} \Delta\tau \leq \epsilon$, whenever $0 < \epsilon < \epsilon_0$. Then the above is bounded by $\epsilon(|h^i(0)| + |F^i|_{L^2})$. Whence,

$$|\begin{pmatrix} I \\ D_\xi \end{pmatrix} \bar{u}^i(\pm L_j^i(\epsilon), \cdot)|_{H^{0.75\times0.25}(\gamma)} \leq \epsilon|h^i(0)| + C|Q_s u_0^i|_{H^1} + C e^{|\gamma|\Delta\tau}|F^i|_{L^2}.$$

The above estimate is also valid in regular layers, with $h^i(0) = 0$. Define,

$$\begin{aligned} \bar{\delta}^i &\stackrel{def}{=} \begin{pmatrix} I \\ D_\xi \end{pmatrix} (\bar{u}^i(L_j^i(\epsilon)) - \bar{u}^{i+1}(-L_j^{i+1}(\epsilon))). \\ |\bar{\delta}^i|_{H^{0.75\times0.25}(I_\tau, \gamma)} &\leq \sup_i \{\epsilon|h^i(0)| + C|Q_s u_0^i|_{H^1} + C e^{|\gamma|\Delta\tau}|F^i|_{L^2}\}. \end{aligned}$$

Consider a new boundary value problem with zero initial condition

$$\bar{\bar{u}}^i_\tau = \mathcal{A}^i \bar{\bar{u}}^i,$$

$$\begin{pmatrix} I \\ D_\xi \end{pmatrix} (\bar{\bar{u}}^i(L^i_j(\epsilon)) - \bar{\bar{u}}^{i+1}(-L^{i+1}_j(\epsilon))) = \bar{\bar{\delta}}^i \stackrel{def}{=} \delta^i - \bar{\delta}^i,$$

$$\bar{\bar{u}}^i = 0.$$

Since $I_\tau$ is a finite interval, $\delta^i$ is also a point in $H^{0.75 \times 0.25}(I_\tau, \gamma)$, with

$$|\delta^i|_{H^{0.75 \times 0.25}(I_\tau, \gamma)} \leq C e^{|\gamma|\Delta\tau} |\delta^i|_{H^{0.75 \times 0.25}(I_\tau)}.$$

$$|\bar{\bar{\delta}}^i|_{H^{0.75 \times 0.25}(I_\tau, \gamma)} \leq C e^{|\gamma|\Delta\tau} |\delta^i|_{H^{0.75 \times 0.25}(I_\tau)}$$
$$+ \sup_i \{\epsilon |\bar{h}^i(0)| + C |Q_s u^i_0|_{H^1} + C e^{|\gamma|\Delta\tau} |F^i|_{L^2}\}.$$

We now use Lemma 4.1, case(2), with $\gamma \leq 0$. If $\epsilon$ is small, the system has a unique solution $\{\bar{\bar{u}}^i\}^\infty_{-\infty}$, with

$$|\bar{\bar{u}}^i|_{H^{2,1}(\Omega^i_j \times I_\tau, \gamma)} \leq C |\{\bar{\bar{\delta}}^i\}|_{H^{0.75 \times 0.25}(\gamma)}.$$

First let $\gamma = 0$ in the above. Combining that with (6.16), and recalling $u^i = \bar{u}^i + \bar{\bar{u}}^i$, we have (6.14).

Now let $-\bar{\sigma}/4 < \gamma < 0$ again. Using a bounded extension of $\bar{\bar{u}}^i$ to $\xi \in \mathbb{R}$. At $\tau = \Delta\tau$, we have

$$|\bar{\bar{h}}^i(\Delta\tau)| + |Q_s \bar{\bar{u}}^i(\Delta\tau)|_{H^1(\mathbb{R})} \leq C |\bar{\bar{u}}^i(\Delta\tau)|_{H^1} \leq C e^{\gamma\Delta\tau} |\bar{\bar{u}}^i|_{H^{2,1}(\Omega^i_j \times I_\tau, \gamma)}$$
$$\leq C (|\{\delta^i\}|_{H^{0.75 \times 0.25}(I_\tau)} + e^{\gamma\Delta\tau} \sup_i \{\epsilon |h^i(0)| + |Q_s u^i_0|_{H^1} + e^{|\gamma|\Delta\tau} |F^i|_{L^2}\}).$$

Combining this with (6.17), and recalling that $u^i = \bar{u}^i + \bar{\bar{u}}^i$, we have

$$|h^i(\Delta\tau)| + |Q_s u^i(\Delta\tau)|_{H^1} \leq C |\{\delta^i\}|_{H^{0.75 \times 0.25}(I_\tau)}$$
$$+ \sup_i \{C_1 e^{\gamma\Delta\tau} |Q_s u^i_0|_{H^1} + (1 + C\epsilon)|h^i(0)| + C e^{\mu\Delta\tau} \sqrt{\Delta\tau} |F^i|_{L^2}\}.$$

Here, we used the fact $|\mu| \leq C\epsilon$ and $e^{\mu\Delta\tau} \leq 1 + e^{\mu\Delta\tau} \mu\Delta\tau$ to simplify (6.17). Observe that $C_1$ does not depend on $\epsilon$ or $\Delta\tau$. Let $\Delta\tau$ be sufficiently large so that $C_1 e^{\gamma\Delta\tau} \leq 1$. From this, estimate (6.15) follows. $\square$

**Proof of Theorem 6.1.** Let the solution of Lemma 6.4 be denoted by

$$\{u^i\} = \mathcal{F}(\{u^i_0\}, \{\delta^i\}, \{F^i\}).$$

Let $U^i \in H^{2,1}(\Omega^i \times I_\tau)$ with $|U^i|_{H^{2,1}} \leq \epsilon$. Let

$$\{\bar{u}^i\} = \mathcal{F}(\{u^i_0\}, \{\delta^i\}, \{\mathcal{N}^i(U^i, \xi, \bar{t} + \epsilon\tau, \epsilon)\}).$$

Based on (4.4), $|\mathcal{N}^i|_{L^2} = o(\epsilon)$. Thus, if $\epsilon$ is small, from Lemma 6.4, $|\bar{u}^i|_{H^{2,1}} \leq \epsilon$. Therefore, $\mathcal{F}$ maps an $\epsilon$-ball in $H^{2,1}$ into itself. From (4.4), it is also clear that $\mathcal{F}$ is a contraction, provided that $\epsilon$ is small. Thus there exits a unique fixed point $\{u^i\}_{-\infty}^{\infty}$ for $\mathcal{F}$. The desired estimate follows from Lemma 6.4 and the fact

$$|\mathcal{N}^i|_{L^2} \leq |g^i|_{L^2} + O(\epsilon)|u^i|_{H^{2,1}}$$

$\square$

**Proof of Theorem 6.2.** Let $\Delta\tau$ be as in Lemma 6.4, and let $I_j$, $0 \leq j \leq r-1$, $I_\infty$ be defined by (6.1) from this $\Delta\tau$.

The spectral projections used in Lemma 6.4 actually depend on the layer index $i$ and the index $j$ of time intervals, and shall be denoted by $Q_0^{i,j}$ and $Q_s^{i,j}$. In regular layers, of course, $Q_0^{i,j} = 0$. Let the critical eigenvalue be $\lambda_j^i(\epsilon)$. The eigenvector $\phi_j^i$ corresponding to $\lambda_j^i(\epsilon)$ and the covector $\psi_j^i$ associated to the adjoint equation are normalized so that $|\phi_j^i|_{L^2} = 1$ and $< \psi_j^i, \phi_j^i >= 1$. The perturbation theory of eigenvalues and eigenvectors yields, in internal layers,

$$|\phi_j^i - \phi_{j+1}^i|_{L^2} + |\psi_j^i - \psi_{j+1}^i|_{L^2} = O(\epsilon). \tag{6.18}$$

Consider the solution of (6.2)–(6.3) as in Theorem 6.1. When $|\{u_j^i(0)\}|_{H^1} = o(\epsilon)$, $|\{\delta_j^i\}|_{H^{0.75 \times 0.25}(I_\tau)} = o(\epsilon)$ and $|\{g_j^i\}|_{L^2} = o(\epsilon)$, the unique solution in that theorem also satisfies $|\{u_j^i\}|_{H^{2,1}} = o(\epsilon)$. Now using Lemma 6.4, with $\mathcal{N}_j^i$ replacing $F^i$, we have

$$|h_j^i(\Delta\tau)| + |Q_s^{i,j}u_j^i(\Delta\tau)|_{H^1(\mathbb{R})} \leq (1 + C\epsilon)\sup_i(|h_j^i(0)| + |Q_s^{i,j}u_j^i(0)|_{H^1})$$
$$+ C|\{\delta_j^i\}|_{H^{0.75 \times 0.25}(I_\tau)} + C|\{\mathcal{N}_j^i\}|_{L^2}.$$

But from (4.4),

$$|\mathcal{N}_j^i|_{L^2} \leq |\{g_j^i\}|_{L^2} + C\epsilon|\{u_j^i\}|_{H^{2,1}}$$
$$\leq |\{g_j^i\}|_{L^2} + C\epsilon(|\{u_j^i(0)\}|_{H^1} + |\{g_j^i\}|_{L^2} + |\{\delta_j^i\}|_{H^{0.75 \times 0.25}(I_\tau)}).$$

Therefore

$$|h_j^i(\Delta\tau)| + |Q_s^{i,j}u_j^i(\Delta\tau)|_{H^1(\mathbb{R})} \leq (1 + C\epsilon)\sup_i(|h_j^i(0)| + |Q_s^{i,j}u_j^i(0)|_{H^1})$$
$$+ C|\{g_j^i\}|_{L^2} + C|\{\delta_j^i\}|_{H^{0.75 \times 0.25}(I_\tau)}.$$

Consider now the initial data in the next time interval $[(j+1)\Delta t, (j+2)\Delta t]$,

$$u_{j+1}^i(\xi, 0) = u_j^i(\xi_1, \Delta\tau) = u_j^i(\nu_3 \circ \nu_2 \circ \nu_1(\xi), \Delta\tau).$$

Cf. (6.4). From the definitions of the mappings $\nu_1$, $\nu_2$, $\nu_3$, we can verify that

$$\xi_1 - \xi = \nu_3 \circ \nu_2 \circ \nu_1(\xi) - \xi = \zeta(\epsilon\xi),$$

where $\zeta$ is a smooth function satisfying

$$|\zeta(\epsilon\xi)| \leq C\epsilon; \quad |\zeta|_{C^1} \leq C. \tag{6.19}$$

The constants C depend on $\Delta\tau$ but not on $\epsilon$. The proof of (6.19) uses the coordinate change formula in §4 and is tedious,. Details will not be given here. Based on (6.19), we can show,

$$|g(\xi_1) - g(\xi)| \leq C\epsilon|\xi||\partial_\xi g|_\infty. \tag{6.20}$$

For any $g \in H^1(\mathbb{R})$, since $\xi \to \xi_1$ is a near identity mapping, we can verify that,

$$|g \circ \nu_3 \circ \nu_2 \circ \nu_1|_{H^1} \leq (1 + C\epsilon)|g|_{H^1}. \tag{6.21}$$

In regular layers, from (6.21), we have, since $Q_0^{i,j+1} = 0$,

$$|Q_s^{i,j+1} u_{j+1}^i(\cdot, 0)|_{H^1} \leq (1 + C\epsilon)|Q_s^{i,j} u_j^i(\cdot, \Delta\tau)|_{H^1}.$$

In internal layers, observe that for $g \in L^2$, $Q_0^{i,j+1} g = (\int_{-\infty}^\infty \psi_{j+1}^i g \, d\xi)\phi_{j+1}^i$. Thus,

$$h_{j+1}^i(0) = \int \psi_{j+1}^i(\xi)[h_j^i(\Delta\tau)\phi_j^i(\xi_1) + Q_s^{i,j} u_j^i(\xi_1, \Delta\tau)]d\xi.$$

$$\begin{aligned}
\int \psi_{j+1}^i(\xi)\phi_j^i(\xi_1)d\xi &= \int \psi_j^i(\xi)\phi_j^i(\xi)d\xi \\
&\quad + \int (\psi_{j+1}^i(\xi) - \psi_j^i(\xi))\phi_j^i(\xi)d\xi + \int \psi_{j+1}^i(\phi_j^i(\xi_1) - \phi_j^i(\xi))d\xi \\
&\leq 1 + C\epsilon.
\end{aligned}$$

In proving the above, (6.18), (6.20) are used.

Observing that the up to the second derivatives of $\phi_j^i(\xi)$ approach 0 exponentially as $\xi \to \pm\infty$, uniformly with respect to $i$, $j$. From (6.19), and the Mean Value Theorem, we have

$$|\phi_j^i(\xi_1) - \phi_j^i(\xi)| + |\partial_\xi(\phi_j^i(\xi_1) - \phi_j^i(\xi))| \leq K\epsilon|\xi|e^{-\gamma_1|\xi|},$$

for some $\gamma_1$, $K > 0$. Therefore,

$$|\phi_j^i \circ \nu_3 \circ \nu_2 \circ \nu_1 - \phi_j^i|_{H^1} \leq C\epsilon.$$

Using the above, evaluating $Q_s^{i,j+1}\phi_j^i = Q_s^{i,j+1}(\phi_j^i - \phi_{j+1}^i)$, then using (6.18), we can prove that

$$|Q_s^{i,j+1}(\phi_j^i \circ \nu_3 \circ \nu_2 \circ \nu_1)|_{H^1} \leq C\epsilon.$$

One can similarly show that,

$$|\int \psi_{j+1}^i(\xi)Q_s^{i,j} u_j^i(\xi_1, \Delta\tau)d\xi| \leq C\epsilon|Q_s^{i,j} u_j^i(\cdot, \Delta\tau)|_{L^2}.$$

$$|Q_s^{i,j+1}(Q_s^{i,j} u_j^i(\nu_3 \circ \nu_2 \circ \nu_1(\cdot), \Delta\tau))|_{H^1} \leq (1 + C\epsilon)|Q_s^{i,j} u_j^i(\cdot, \Delta\tau)|_{H^1}.$$

Combining all these estimates, we have shown,

$$\begin{pmatrix} |h^i_{j+1}(0)| \\ |Q^{i,j+1}_s u^i_{j+1}(0)|_{H^1} \end{pmatrix} \le \begin{pmatrix} 1+C\epsilon & C\epsilon \\ C\epsilon & (1+C\epsilon) \end{pmatrix} \begin{pmatrix} |h^i_j(\Delta\tau)| \\ |Q^{i,j}_s u^i_j(\Delta\tau)|_{H^1} \end{pmatrix}.$$

Therefore,

$$|h^i_{j+1}(0)| + |Q^{i,j+1}_s u^i_{j+1}(0)|_{H^1} \le (1+C\epsilon)(|h^i_j(\Delta\tau)| + |Q^{i,j}_s u^i_j(\Delta\tau)|_{H^1}).$$

Combining the above with Lemma 6.4, we have

$$|h^i_{j+1}(0)| + |Q^{i,j+1}_s u^i_{j+1}(0)|_{H^1}$$
$$\le (1+C\epsilon)\sup_i\{|h^i_j(0)| + |Q^{i,j}_s u^i_j(0)|_{H^1}\} + C(|\{\delta^i_j\}|_{H^{0.75\times0.25}} + |\{g^i_j\}|_{L^2}).$$

By induction, the initial condition $\{u^i_j(0)\}$ for the $j$-th time interval satisfies

$$|h^i_j(0)| + |Q^{i,j}_s u^i_j(0)|_{H^1} \le (1+C\epsilon)^j\sup_i\{|h^i_0(0)| + |Q^{i,0}_s u^i_0|_{H^1}\}$$
$$+ \sum_{k=0}^{j-1}(1+C\epsilon)^{j-k-1}C(|\{\delta^i_k\}|_{H^{0.75\times0.25}} + |\{g^i_k\}|_{L^2})$$
$$\le (1+C\epsilon)^j\sup_i\{|h^i_0(0)| + |Q^{i,0}_s u^i_0|_{H^1}\}$$
$$+ \frac{(1+C\epsilon)^j}{\epsilon}(|\{\delta^i\}|_{H^{0.75\times0.25}} + |\{g^i\}|_{L^2}).$$

Since $j \le r \le \log(1/(C_0\epsilon^2))/(\epsilon\bar\gamma\Delta\tau) + 1$,

$$(1+C\epsilon)^j \le (1+C\epsilon)(1+C\epsilon)^{\frac{1}{C\epsilon}\log(1/(C_0\epsilon^2))\frac{C}{\bar\gamma\Delta\tau}}$$
$$\le (1+C\epsilon)e^{\log(1/(C_0\epsilon^2))\frac{C}{\bar\gamma\Delta\tau}}$$
$$\le (1+C\epsilon)[C_0\epsilon^2]^{\frac{-C}{\bar\gamma\Delta\tau}} \le C\epsilon^{-\frac{2C}{\bar\gamma\Delta\tau}}.$$

The estimate in Theorem 6.2 follows easily.                            □

## 7. Application to Singularly Perturbed Reaction Diffusion Systems

We first review the construction of matched asymptotic expansions for a wave-front-like solution for system (1.1), cf. [25]. We then show that the *Spatial Shadowing Lemma* can be used to obtain exact solutions near the formal expansion.

As in §1, assume $f(u,x,\epsilon) = \sum_{j=0}^{\infty}\epsilon^j f_j(u,x)$, and $f(u,x+x_p,\epsilon) = f(u,x,\epsilon)$ for some $x_p > 0$. We will consider internal and regular layers occurring alternatively in the $x$ direction.

Assume there is a partition $\mathbb{R} = \cup_{\ell=-\infty}^{\infty}[x^\ell, x^{\ell+1}]$ that is periodic with respect to $\ell$, compatible with the period of $f$. That is, there exists $\ell_p$ such that $x_{\ell+\ell_p} = x_\ell + x_p$. A $C^\infty$ function $p^i(x)$ is defined on $[x^{i-1}, x^i]$ with $f_0(p^i(x), x) = 0$. Also assume that $p^{\ell+\ell_p}(x + x_p) = p^\ell(x)$, $x \in [x^{\ell-1}, x^\ell]$.

**H* 1** $\mathrm{Re}\,\sigma\{f_{0u}(p^i(x), x)\} < 0$ for $x \in [x^{i-1}, x^i]$, $i \in \mathbb{Z}$.

Using **H*1**, a formal expansion

$$\sum_0^\infty \epsilon^j u_j^{Ri}(x), \quad u_0^{Ri} = p^i(x),$$

for (1.1) can easily be obtained. Note that the expansion is time-independent, since f is so.

Using a stretched variable $\xi = \dfrac{x - x^i}{\epsilon}$, we assume that the 0th order expansion of Eq. (1.1),

$$u_{\xi\xi} + f_0(u, x^i) = 0. \tag{7.1}$$

has a heteroclinic solution $q^i(\xi)$ connecting $p^i(x^i)$ to $p^{i+1}(x^i)$. Assume

**H* 2** The linear homogeneous equation

$$\phi_{\xi\xi} + f_{0u}(q^i(\xi), x^i)\phi = 0,$$

has a unique bounded solution $q_\xi^i(\xi)$, up to constant multiples.

From **H*2**, we infer that the adjoint equation

$$\psi_{\xi\xi} + f_{0u}^\tau(q_\xi^i(\xi), x^i)\psi = 0$$

has a unique bounded solution $\psi_i(\xi)$ up to constant multiples. Moreover, $\psi(\xi) \to 0$ exponentially as $|\xi| \to \infty$.

The following assumption implies that the heteroclinic solution breaks as $x$ moves away from $x^i$.

**H* 3** $\int_{-\infty}^\infty \psi_i^\tau(\xi) f_{0x}(q^i(\xi), x^i) d\xi \neq 0, \quad i \in \mathbb{Z}$.

Under the hypotheses **H*1** to **H*3**, we can construct formal expansions for the stationary wave front positions and solutions in internal layers

$$\sum_0^\infty \epsilon^j x_j^i, \quad x_0^i = x^i,$$

$$\sum_0^\infty \epsilon^j u_j^{Si}(\xi), \quad u_0^{Si} = q^i.$$

The formal solutions in the internal and regular layers match at their common boundaries. See [23, 25].

When $x$ is in a neighborhood of $x^i$, we look for a travelling wave solution with wave speed $V^i(x)$. It is also of interest to find out conditions to ensure the wave front to move towards $x^i$. Consider $\mathcal{A}^i : L^2(\mathbb{R}) \to L^2(\mathbb{R})$, defined as

$$\mathcal{A}^i u = u_{\xi\xi} + f_{0u}(q^i(\xi), x^i)u, \quad D(\mathcal{A}^i) = H^2(\mathbb{R}).$$

**H* 4** $\lambda = 0$ is a simple eigenvalue for $\mathcal{A}^i$, $i \in \mathbb{Z}$. There exists $\bar{\sigma} > 0$ such that

$$\sigma(\mathcal{A}^i) \subset \{\lambda = 0\} \cup \{\mathrm{Re}\lambda \leq -\bar{\sigma}\}.$$

Hypothesis **H*4** implies that $q^i(\xi)$ is a stable equilibrium for

$$u_\tau = u_{\xi\xi} + f_0(u, x^i), \tag{7.2}$$

modulo a phase shift in $\xi$, see [9, 10, 11, 12]. Since $\lambda = 0$ is simple, we have

$$\int_{-\infty}^{\infty} \psi_i^\tau(\xi) q_\xi^i(\xi) d\xi \neq 0. \tag{7.3}$$

**H* 5** $I(x^i) = [\int_{-\infty}^{\infty} \psi_i^\tau(\xi) q_\xi^i(\xi) d\xi]^{-1} \int_{-\infty}^{\infty} \psi_i^\tau(\xi) f_{0x}(q^i(\xi), x^i) d\xi > 0$, $i \in \mathbb{Z}$.

Condition **H*5** implies that the wave front is moving towards $x^i$, if $x$ is near $x^i$. In fact, it is shown in [25],

$$\frac{\partial V^i(x^i)}{\partial x} = -I(x^i).$$

An important consequence of (7.3) is that if $x$ is near $x^i$, there exists a unique $V = V^i(x)$ such that (7.2) has a travelling wave solution $q^i(\xi, x)$, with the wave speed $V$, satisfying

$$u_{\xi\xi} + V^i(x)u_\xi + f_0(u, x) = 0. \tag{7.4}$$

The function $q^i(\xi, x)$ connects $p^i(x)$ and $p^{i+1}(x)$ as $\xi \to \pm\infty$. More precisely, under the Hypotheses **H*1** to **H*5**, there exists an open interval $O^i$, containing $x^i$ such that the following properties holds. For a proof, see [25], Lemma 4.1.

(i) $p^i(x)$ and $p^{i+1}(x)$ can both be extended smoothly to $O^i$ with

$$f_0(p^j(x), x) = 0,$$
$$\mathrm{Re}\sigma(f_{0u}(p^j(x), x)) < 0, \quad j = i, \, i+1.$$

(ii) There exists a $C^\infty$ function $V^i : O^i \to \mathbb{R}$, such that for each $x \in O^i$, Eq. (7.2) has a unique heteroclinic solution $q^i(\xi, x)$, connecting $p^i(x)$ to $p^{i+1}(x)$, with $q^i(0, x) - q^i(0) \perp q_\xi^i(0)$. In particular, $V^i(x^i) = 0$ and $q^i(\xi, x^i) = q^i(\xi)$. Moreover $q^i(\xi, x)$ is $C^\infty$ in both variables.

(iii) The linear equation

$$\phi_{\xi\xi} + V^i(x)\phi_\xi + f_{0u}(q^i(\xi, x^i), x)\phi = 0,$$

has a unique bounded solution $q_\xi^i(\xi, x)$ up to constant multiples. The adjoint equation

$$\psi_{\xi\xi} - v^i(x)\psi_\xi + f_{0u}^\tau(q^i(\xi, x^i), x)\psi = 0,$$

has a unique bounded solution $\psi_i(\xi, x)$, normalized so that

$$\int_{-\infty}^{\infty} \psi_i(\xi, x)q_\xi^i(\xi, x)d\xi = 1.$$

$\psi(\cdot, x)$ is a $C^\infty$ function of $x$ in the space $C(\mathbb{R})$.

(iv) The densely defined closed operator $\mathcal{A}^i(x) : L^2 \to L^2$,

$$\mathcal{A}^i(x)u = u_{\xi\xi} + V^i(x)u_\xi + f_{0u}(q^i(\xi, x), x)u,$$

has $\lambda = 0$ as a simple eigenvalue with eigenvector $q_\xi^i(\cdot, x)$. All the rest of the spectrum are contained in $\text{Re}\,\lambda \le -\alpha_0$ for some $\alpha_0 > 0$.

(v) For $x \in O^i$, $i \in \mathbb{Z}$,

$$I(x) = \int_{-\infty}^{\infty} \psi_i^\tau(\xi, x)f_{0u}(q^i(\xi, x), x)d\xi > 0,$$

As a consequence of (v), $V^i(x) > 0$, $(< 0)$ if $x < x^i$, $(> x^i)$, and

$$\frac{\partial V^i(x)}{\partial x} = -I(x), \ x \in O^i, \ i \in \mathbb{Z}.$$

Under the Hypotheses $\mathbf{H^*1}$ to $\mathbf{H^*5}$, we can construct formal series solutions and wave positions for (1.1) in the $i$-th internal layer,

$$\sum_0^\infty \epsilon^j u_j^{Si}(\xi, t), \quad \sum_0^\infty \epsilon^j \eta_j^i(t).$$

The formal expansions are to be truncated to the $m$th order, as in (1.2), where $u_j^{R\ell}(x, t) = u_j^{R\ell}(x)$ is, in fact, $t$-independent. Based on these expansions, we define a formal approximation $\{w^i\}$ by (1.5). In the rest of the section, we show that Hypotheses $\mathbf{H1}$–$\mathbf{H4}$ in §2 are satisfied by this $\{w^i\}$. Besides, by choosing large $m$, the results in [25] indicate that $j_2$, as in (2.4), can be arbitrarily large. Therefore, the existence of exact solutions near $\{w^i\}$ is ensured by the *Spatial Shadowing Lemma* as in §2.

To the zero-th order, the wave position obeys the equation

$$\frac{dx}{dt} = V^i(x),$$

with an initial condition $x(0) = \eta_0^i(0) \in O^i$. $x = x^i$ is a stable equilibrium due to the condition $D_x V^i(x^i) < 0$. Therefore

$$|\eta_0^i(t) - x^i| \le Ce^{-\bar{\gamma}t},$$

for some $C$, $\bar{\gamma} > 0$. In fact, it is shown in [25] that

$$|D_t \eta_j^i(t)| + |\eta_j^i(t) - \eta_j^i(\infty)| \leq C_j e^{-\bar{\gamma}t}, \quad j \in \mathbb{Z},$$

for some $C_j > 0$. Moreover, in the weighted norm,

$$\sup_\xi \{(1 + |\xi|^j)^{-1} \, | \, u_j^{Si}(\xi, t) - u_j^{Si}(\xi, \infty)\} \leq C_j e^{-\bar{\gamma}t},$$

for some $j \geq 0$. In the domain $|\xi| \leq \epsilon^{\beta-1}$, we have $(\epsilon|\xi|)^j \ll 1$, and

$$\sum_{j=0}^m \epsilon^j |u_j^{Si}(\xi, \tau) - u_j^{Si}(\xi, \infty)| \leq \bar{C}_m e^{-\bar{\gamma}t}. \qquad (7.5)$$

Also

$$\sum_{j=0}^m \epsilon^j |D_t \eta_j^i(t)| + \sum_{j=0}^m \epsilon^j |\eta_j^i(t) - \eta_j^i(\infty)| \leq \bar{C}_m e^{-\bar{\gamma}t}. \qquad (7.6)$$

The constant $C_m$ depends on the order of truncation $m$. It follows from (7.5) and (7.6) that **H1** in §1 is satisfied in internal layers. **H1** is certainly satisfied in regular layers where the expansions are $t$-independent.

From **H*1**, the Hypothesis **H2** in §2 is satisfied if $w^i = p^i(x)$. In general, adding higher order terms introduces a perturbation of $O(\epsilon)$. From the perturbation theory of eigenvalues of the matrix $f_{0u}$, **H2** in §2 is valid if $w^i$ is an $m$-th order truncation of the formal solution.

Let $\sum_0^\infty \tilde{\epsilon}^j u_j^{Ri}(\xi, t)$ be the inner expansion of the outer solution $\sum \epsilon^j u_j^{Ri}(x, t)$, [25, 7, 8]. For any $\mu > 0$, if $\epsilon$ is small, we have, in the region $|\xi| \leq \epsilon^{\beta-1}$,

$$|\sum_0^m \epsilon^j u_j^{Ri}(\eta^i(t, \epsilon) + \epsilon\xi, t) - \sum_0^m \tilde{u}_j^{Ri}(\xi, t)| = O(\epsilon^{m+1}|\xi|^{m+1})$$

$$= O(\epsilon^{\beta(m+1)}) \leq \mu/3.$$

It is shown in [25], based on the matching of expansions in internal and regular layers, there exists $\gamma_1 > 0$ such that

$$\sum_0^m \epsilon^j |u_j^{Si}(\xi, t) - \tilde{u}_j^{Ri}(\xi, t)| \leq C \sum_0^m \epsilon^j (1 + |\xi|^j) e^{-\gamma_1|\xi|}.$$

In the domain $|\xi| \leq \epsilon^{\beta-1}$, $\epsilon^j(1 + |\xi|^j) \ll 1$, thus the above is bounded by $Ce^{-\gamma_1|\xi|}$. For $\mu > 0$, there exists $N > 0$ such that the above is bounded by $\mu/3$ for $N \leq \xi \leq \epsilon^{\beta-1}$. Observe that when $|\xi| \leq \epsilon^{\beta-1}$, and $\epsilon$ is small,

$$|\sum_0^m \epsilon^j u_j^{Ri}(\eta^i(t, \epsilon) + \epsilon^\beta, t) - \sum_0^m \epsilon^j u_j^{Ri}(\eta^i(t, \epsilon) + \epsilon\xi, t)|$$

$$\leq C|\epsilon^\beta - \epsilon\xi| < \mu/3.$$

Combining the three estimates, we have proved **H3**, §2 for $\epsilon^{\beta-1} \geq \xi \geq N$. The proof for $-\epsilon^{\beta-1} \leq \xi \leq -N$ is similar.

It remains to show **H4** of §2. From the Property (iv) of $O^i$, when $\epsilon = 0$, $\mathcal{A}^i(t)$ has a simple eigenvalue $\lambda = 0$, all the other spectra are contained in $\text{Re}\lambda \leq -\sigma_0$. By a standard perturbation analysis, for any $\delta > 0$, there exists $\epsilon_0$ such that for $0 < \epsilon < \epsilon_0$, $\mathcal{A}^i(t)$ has a simple eigenvalue $|\lambda^i(\epsilon)| < \delta$, while all the other spectra are contained in $\text{Re}\lambda \leq -\sigma_0 + \delta$. Moreover, the eigenvector corresponding to the critical eigenvalue can be written as $q_\xi^i + z$, with $z(0) \perp q_\xi^i(0)$ and $|z| < \delta$. it is also clear that $\lambda^i(t, \epsilon) = \epsilon \lambda_0^i(t) + O(\epsilon^2)$ due to the smooth dependence of eigenvalue on $\epsilon$. Details will not be rendered here. We now show the following important result.

**Lemma 7.1.**

$$\lambda_0^i(t) = \frac{\partial \lambda^i(\epsilon)}{\partial \epsilon} = \frac{\partial V^i(x)}{\partial x} + [\int_{-\infty}^{\infty} \psi_i^\tau(\xi) q_\xi^i(\xi) d\xi]^{-1} [\int_{-\infty}^{\infty} \psi_i^\tau(\xi) q_{x\xi}^i(\xi) d\xi] V^i(x).$$

*Proof.* The eigenvalue $\lambda$ and eigenvector $z + q_\xi^i$ satisfy

$$\lambda(z + q_\xi^i) = (z + q_\xi^i)_{\xi\xi} + D_t \eta^i(t, \epsilon)(z + q_\xi^i)_\xi + f_u(u^i, \eta^i + \epsilon\xi, \epsilon)(z + q_\xi^i).$$

Here $u^i = \sum_0^m \epsilon^j u_j^i$ and $\eta^i = \sum_0^m \epsilon^j \eta_j^i$ are formal expansions for the solutions and wave front in the $i$th internal layer. Using the fact $q_\xi^i$ is an eigenvector corresponding to $\lambda = 0$, we rewrite the equation as a nonhomogeneous equation for $z$ with forcing term that will be zero if $\epsilon = \lambda = 0$. The operator

$$z \to z_{\xi\xi} + D_t \eta_0^i(t) z_\xi + f_{0u}(q^i(\xi), x), x)z$$

is Fredholm, in the space $L^2(|\xi|) = \{u : u(\xi)/(1+|\xi|) \in L^2\}$ of weighted $L^2$ functions, with one dimensional kernel and co-kernel. The reason to use weighted norm is that the partial derivative of the right hand side with respect to $\epsilon$ yields a factor $\xi$. Lyapunov-Schmidt method can be used to deduce that there exists a unique solution $z$, $z(0) \perp q_\xi^i(0)$ if a bifurcation equation $G(\lambda, \epsilon) = 0$ is satisfied. The partial derivatives can be obtained as Melnikov type integrals.

$$\frac{\partial G(0,0)}{\partial \lambda} = -\int_{-\infty}^{\infty} \psi_i^\tau(\xi) q_\xi^i(\xi) d\xi.$$

$$\frac{\partial G(0,0)}{\partial \epsilon} = \int_{-\infty}^{\infty} \psi_i^\tau(\xi) \{D_t \eta_1^i(t) q_{\xi\xi}^i + f_{0uu} u_1^i q_\xi^i + f_{0ux} \cdot (\eta_1^i(t) + \xi) q_\xi^i + f_{1u} \cdot q_\xi^i\} d\xi.$$

Introducing a variable $\bar{\xi}$, we can rewrite the above as, $(x = \eta_0^i(t))$,

$$\frac{\partial G(0,0)}{\partial \epsilon} = \int_{-\infty}^{\infty} \psi_i^\tau(\xi + \bar{\xi}) D_\xi \{D_t \eta_1^i(t) q_\xi^i(\xi + \bar{\xi}) + f_{0u}(q^i(\xi + \bar{\xi}, x), x) u_1^i(\xi + \bar{\xi})$$
$$+ f_{0x}(q^i(\xi + \bar{\xi}, x), x)(\eta_1^i(t) + \xi + \bar{\xi}) + f_1(q^i(\xi + \bar{\xi}, x), x)\} d\xi$$
$$- \int_{-\infty}^{\infty} \psi_i^\tau(\xi + \bar{\xi})[f_{0u}(q^i(\xi + \bar{\xi}, x), x) u_{1\xi}^i(\xi + \bar{\xi}) + f_{0x}(q^i(\xi + \bar{\xi}, x), x)] d\xi$$

Observe that $u_0^i = q^i$, and

$$u_{0t}^i = u_{1\xi\xi}^i + D_t \eta_0^i(t) u_{1\xi}^i + f_{0u} u_1^i + D_t \eta_1^i(t) u_{0\xi}^i + f_{0x} \cdot (\eta_1^i(t) + \xi) + f_1,$$

which can be obtained from expanding (1.1) in powers of $\epsilon$, see [25]. We have,

$$\frac{\partial G}{\partial \epsilon} = \int_{-\infty}^{\infty} \psi_i^\tau(\xi) \{D_\xi[-u_{1\xi\xi}^i - D_t \eta_0^i(t) u_{1\xi}^i + u_{0t}^i] - f_{0u} \cdot u_{1\xi}^i - f_{0x}\} d\xi.$$

Observe that $H = (u_{1\xi}^i)_{\xi\xi} + D_t \eta_0^i(t)(u_{1\xi}^i)_\xi + f_{0u} u_{1\xi}^i$ is in the range of the operator $\mathcal{A}^i$ (when $\epsilon = 0$). Thus $\int_{-\infty}^{\infty} \psi_i^\tau H d\xi = 0$. Also observe $u_{0t}^i = D_t q^i(\xi, \eta^i(t)) = q_x^i V^i(x)$. Therefore, we have

$$\frac{\partial G}{\partial \epsilon} = \int_{-\infty}^{\infty} \psi_i^\tau(\xi) \{q_{x\xi}^i(\xi, x) V^i(x) - f_{0x}\} d\xi.$$
$$\frac{\partial \lambda}{\partial \epsilon} = [\int_{-\infty}^{\infty} \psi_i^\tau q_\xi^i d\xi]^{-1} \{[\int_{-\infty}^{\infty} \psi_i^\tau q_{x\xi}^i d\xi] V^i(x) - \int_{-\infty}^{\infty} \psi_i^\tau f_{0x} d\xi\}$$

The desired result follows from the following formula, see [25],

$$\frac{\partial V^i(x)}{\partial x} = -[\int_{-\infty}^{\infty} \psi_i^\tau q_\xi^i d\xi]^{-1} [\int_{-\infty}^{\infty} \psi_i^\tau f_{0x} d\xi].$$

$\square$

**Corollary 7.2.** *If $\frac{\partial V^i(x^i)}{\partial x} \neq 0$, then there exists a neighborhood of $x^i$ where*

$$sign(\lambda_0^i(t)) = sign(\frac{\partial V^i(x^i)}{\partial x})$$

The proof of Corollary 7.2 is based on $V^i(x^i) = 0$, and $V^i(x)$ is small if $x$ is near $x^i$. From $\mathbf{H}^*\mathbf{5}$, it is clear that $\lambda_0^i(\infty) < 0$, since $\eta_0^i(\infty) = x^i$. The method of introducing a new variable $\bar{\xi}$ in the proof of Lemma 7.1 has been used in [23].

Department of Mathematics, North Carolina State University, Raleigh, North Carolina 27695–8205

*E-mail address*: xblin@xblsun.math.ncsu.edu